\documentclass[a4paper,12pt,english]{article}

\usepackage{amsfonts,bm,amssymb,euscript,array,babel,cite}

 \usepackage[all]{xy}

\usepackage{amsmath,amsthm,amscd}

\usepackage[dvips]{epsfig}

\usepackage{slashed}

\usepackage[ansinew]{inputenc}



\newcommand{\be}{\begin{equation}}
\newcommand{\ee}{\end{equation}}

\newcommand{\beq}{\begin{equation}}

\newcommand{\eeq}{\end{equation}}

\newcommand{\beqa}{\begin{eqnarray}}

\newcommand{\eeqa}{\end{eqnarray}} 

\def\nn{\nonumber} \def \bea{\begin{eqnarray}} \def\eea{\end{eqnarray}}

\newcommand{\barr}{\begin{array}}

\newcommand{\earr}{\end{array}}

\numberwithin{equation}{section}

\def\a{\alpha}  

 \def\g{\gamma} \def\G{\Gamma}

 \def\d{\delta}

\def\mc{\mathcal}

\def\R{{\mathbb R}}  

\def\Z{{\mathbb Z}} \def\one{\mbox{1 \kern-.59em {\rm l}}}

\def\bit{\begin{itemize}} \def\eit{\end{itemize}}

\def\({\left(} \def\){\right)}

 \allowdisplaybreaks[3]

\begin{document}

\makeatother


\parindent=0cm

\renewcommand{\title}[1]{\vspace{10mm}\noindent{\Large{\bf

#1}}\vspace{8mm}} \newcommand{\authors}[1]{\noindent{\large

#1}\vspace{5mm}} \newcommand{\address}[1]{{\itshape #1\vspace{2mm}}}

\begin{titlepage}

\begin{center}

\title{ \Large Matrix theory compactifications on twisted tori}

\vskip 3mm

\authors{Athanasios {Chatzistavrakidis${}^{1}$} and Larisa {Jonke${}^{1,2}$}}

\vskip 3mm

\address{

{${}^1$}{\it Bethe Center for Theoretical Physics and Physikalisches Institut, University of Bonn, \\

Nussallee 12, D-53115 Bonn, Germany}  }

\bigskip 

\address{{${}^2$}{\it Theoretical Physics Division, Rudjer Bo\v skovi\'c Institute, \\~Bijeni\v cka 54, 10000  Zagreb, Croatia} 
}

\bigskip

E-mails:
than@th.physik.uni-bonn.de, larisa@irb.hr

\date{}

\vskip 1.4cm

\textbf{Abstract}

\vskip 3mm

\begin{minipage}{14cm}%

\end{minipage}

\end{center}

We study compactifications of Matrix theory on twisted tori and non-commutative versions of them. 
As a first step, we review the construction of multidimensional twisted tori realized as nilmanifolds 
based on certain nilpotent Lie algebras. Subsequently, matrix compactifications on tori are revisited 
and the previously known results are supplemented with a background of a non-commutative torus with 
non-constant non-commutativity and an underlying non-associative structure on its phase space. 
Next we turn our attention to 3- and 6-dimensional twisted tori and we describe consistent backgrounds 
of Matrix theory on them by stating and solving the conditions which describe the corresponding 
compactification. Both commutative and non-commutative solutions are found in all cases. Finally, 
we comment on the correspondence among the obtained solutions and flux compactifications of 
11-dimensional supergravity, as well as on relations among themselves, such as Seiberg-Witten maps 
and T-duality.

\end{titlepage}

\tableofcontents


\section{Introduction}

An attractive way to gain access to the non-perturbative regime of superstring theories and M theory passes through 
certain matrix models. The most prominent instances include the matrix model of 
Banks-Fischler-Shenker-Susskind (BFSS) \cite{Banks:1996vh}, also known as Matrix theory, and the type IIB matrix model of 
Ishibashi-Kawai-Kitazawa-Tsuchiya (IKKT) \cite{Ishibashi:1996xs}. They were suggested as 
 non-perturbative definitions of M theory and type IIB superstring theory respectively. As such 
they provide frameworks where  brane dynamics can be studied and non-perturbative duality 
symmetries can be tested.

An interesting program in the study of matrix models, already initiated in \cite{Banks:1996vh} 
and \cite{Taylor:1996ik}, relates to 
their toroidal compactification. This is defined by a restriction of the action functional of the 
model under certain conditions incorporating the geometry of the 
compactification space.
In \cite{cds}, Connes, Douglas and Schwarz perform a detailed study of 
matrix compactifications on multidimensional tori and unveil striking relations to non-commutative 
geometry. They argue that non-commutative deformations of tori are tantamount to turning on fluxes in 
supergravity compactifications. Such a correspondence was supported by subsequent work on the subject 
\cite{Kawano:1998re,Douglas:1997fm,Brace:1998ku,ramgoolam}. Matrix compactifications on spaces other than tori were considered 
in \cite{Ho:1997yk,Ho:1998xh}. 

In this paper we revisit matrix compactifications on tori and perform a study of compactifications on multidmensional 
twisted tori. The latter are smooth manifolds corresponding to non-trivial fiber bundles with a toroidal 
fiber over a toroidal base. They may be described equivalently as nilmanifolds, namely coset spaces 
obtained by quotiening an appropriate discrete group out of a nilpotent Lie group\footnote{The relevance of non-semisimple Lie groups and their algebras for matrix models was pointed out in \cite{Chatzistavrakidislie} and found an interesting application in \cite{Nishimura}.}. Such a description, 
including illuminating examples, is provided in section 2. 
 In order to set the stage for the compactification of Matrix theory, we review in section 3 
the basics of the BFSS and IKKT matrix models, as well as their toroidal compactification. Apart from the 
previously obtained results, we describe a solution of the associated conditions which corresponds 
to a non-commutative deformation of the torus with non-constant non-commutativity. This solution carries 
an underlying non-associative structure on the corresponding phase space.

The approach of matrix compactifications on tori is utilized, with the necessary modifications, in order to 
study compactifications on twisted tori. An analysis of the case of the twisted 3-torus is performed, 
where a set of solutions to the corresponding conditions is identified for commutative and 
non-commutative twisted 3-tori. The solution already found in \cite{ramgoolam} is recovered too. A similar 
analysis is carried out for a particular 6-dimensional twisted torus. The resulting solutions are presented 
in a form which may be directly generalized to any other higher-dimensional nilmanifold.

The solutions of Matrix theory on tori and twisted tori are related in certain ways among themselves as well 
as with flux compactifications of  string / M theory in the supergravity approximation. Adopting 
the Connes-Douglas-Schwarz correspondence between non-commutative deformations and supergravity fluxes, 
we suggest that the toroidal background with non-constant non-commutativity may be associated with 
a constant 4-form flux in 11-dimensional supergravity. Likewise, non-commutative twisted tori may be 
related to compactifications where both NS fluxes and geometric fluxes are present \cite{km,Hull:2005hk,Hull:2006tp}. Moreover, in section 4 we present certain transformations relating 
the non-commutative solutions  to the commutative ones and inducing the Seiberg-Witten maps \cite{SW}  between the corresponding gauge theories. Finally we comment on the T-duality between toroidal backgrounds 
with non-constant B field and NS-fluxless twisted tori. Section 5 contains our conclusions and 
appendix A contains the geometric data of a certain class of interesting 5- and 6-dimensional nilmanifolds.

\section{Twisted tori as nilmanifolds}

\subsection{General considerations}

In order to generate masses in four dimensions induced by a higher-dimensional supergravity, Scherk and 
Schwarz introduced a general consistent compactification scheme in their pioneering paper \cite{Scherk}. 
Specific realizations of this scheme are provided by the so-called twisted tori and the low-energy 
effective action resulting from a dimensional reduction of the heterotic string on them was studied in 
detail in \cite{km}. Later it was shown that higher-dimensional twisted tori also provide consistent 
backgrounds of type II string theories \cite{kstt,Grana} and M theory \cite{Hull:2005hk,Hull:2006tp}.

Twisted tori may be described in essentially two complementary ways. First of all they arise as T-duals 
of square tori endowed with a constant NS 3-form flux{\footnote{Clearly this corresponds to a case of non-constant B field.}} \cite{kstt}. Here we shall concentrate on the second description, which is 
directly related to the study of nilmanifolds.

Nilmanifolds are smooth manifolds constructed as quotients of nilpotent Lie groups by discrete 
subgroups of them \cite{malcev}. Thus a nilmanifold $\mathcal M$ may be described as a coset space $A/\G $ of a 
nilpotent group $A$ and a discrete group $\G\subset A$. The nilpotent Lie groups of dimension up to six 
and their corresponding Lie algebras $\mathcal A$ are fully classified (see e.g. \cite{Patera:1976ud}). 
Here we shall follow the notation appearing in the tables of \cite{Patera:1976ud}, denoting a nilpotent 
Lie algebra as ${\cal A}_{d,i}$, where $d$ is the dimension of the algebra (the number of its generators) 
and $i$ is just an enumerating index according to the aforementioned tables. Moreover, when there is some 
parameter on which the algebra depends, 
it will appear as superscript, e.g. ${\cal A}_{d,i}^{\a}$ if there is one parameter $\a$. Let us also note 
that the 
generators of an algebra will be denoted as $X_a,a=1,\dots,d$.

Since we are willing to use nilmanifolds for compactification, it is self-evident that they had better be compact manifolds. It turns out that a necessary condition for compactness is that the 
group $A$ is unimodular, i.e. its structure constants satisfy $f^a_{\ ab}=0$ (this was already discussed in
 \cite{Scherk}). This condition is in general not sufficient 
but for nilpotent groups it is enough to require that their structure constants are rational \cite{malcev}. 
Hereforth we rely on the above assumptions.

A fact which facilitates the coset construction of nilmanifolds is that the group elements $g\in A$ may always 
be expressed as upper triangular matrices. Then the algorithm for the construction of a nilmanifold 
consists of the following steps:

\begin{enumerate}

 \item Express the basis elements $X_a$ of the nilpotent Lie algebra $\mathcal A_{d,i}$ as upper triangular 
$d\times d$ matrices.
\item Choose a representative general group element $g\in A_{d,i}$. A convenient choice is: $g=\prod_{a=1}^d\mbox{exp}(x^aX_a),~ x^a\in\R $. 
\item Consider the general discrete subgroup element $\g\in\G$ as the restriction of $g \in A_{d,i}$ for 
integer coefficients: $\g=\prod_{a=1}^d\mbox{exp}(\g^a X_a),~ \g^a\in\Z $.
\item The subgroup $\G $ acts on $A_{d,i}$ by matrix multiplication. Thus as a final step one can 
construct the left coset $\mathcal M=A_{d,i}/\G $.

\end{enumerate}

Having constructed the nilmanifold as above, it is then easy to study its geometry. Most importantly, 
it is straightforward to compute the Lie algebra 1-form
\beq
e=g^{-1}dg, 
\eeq
which is Lie algebra valued and therefore it might be decomposed as $e=e^aX_a$. The quantities $e^a$ 
correspond to the usual vielbein basis and they may be expressed in terms of the coordinate basis 1-forms $dx^a$ as
\be 
e^a=U(x)^a_{b}dx^{b},
\ee
for some $x$-dependent twist matrix $U(x)$. The vielbeins satisfy the Maurer-Cartan equations 
\be de^a=-\frac 12 f_{bc}^ae^b\wedge e^c, \ee 
 where $f_{bc}^a$ are constant coefficients, which are identified with the structure constants of the 
Lie algebra $\mathcal A_{d,i}$. In the context of flux compactifications they are also referred to as 
geometric fluxes.

Finally, in accord with the above, it is very useful to read off the coordinate identifications which 
are made in the process of the compactification of the nilpotent group. For the square torus, e.g. in 
three dimensions with coordinates $x^1,x^2,x^3$ and unit radii, these identifications are very simple and they read 
as 
\be \label{t3id}
(x^1,x^2,x^3)\sim (x^1+1,x^2,x^3)\sim (x^1,x^2+1,x^3)\sim (x^1,x^2,x^3+1).
\ee  
The corresponding identifications for nilmanifolds are slightly less simple but they are easily obtained 
from the twist matrix $U(x)$, as it will become evident in the following. For example, in the case of a 
3-dimensional nilmanifold with coordinates $x^a,~a=1,2,3$ and unit radii, we shall show that they are
\be (x^1,x^2,x^3)\sim (x^1+1,x^2,x^3)\sim(x^1,x^2,x^3+1)\sim(x^1+x^3,x^2+1,x^3).\ee 
A very illuminating consequence is that the latter identifications allow the interpretation of the 
3-dimensional nilmanifold as twisted fibration of a 2-torus over a circle, namely a 2-torus in the $x^1,x^3$ 
directions whose geometry varies as it traverses the base circle in the $x^2$ direction. This is the 
reason why we may call it a twisted torus. In fact, the above observation holds  in higher dimensions as 
well, thus providing a correspondence between nilmanifolds and twisted fibrations of toroidal fibers over 
toroidal bases.

\subsection{A 3-dimensional example}

Let us now provide a couple of examples of the previously described procedure. We start with the simplest 
possible case, based on the unique 3-dimensional nilpotent Lie algebra, which is the algebra of the Weyl 
group. We use the notation $\mathcal A_{3,1}$, in accord with \cite{Patera:1976ud}. The only non-trivial commutation relation of this algebra is 
\be\label{cr31} [X_2,X_3]=X_1. \ee  
Following the steps which were described in the previous subsection, first we write down a basis 
for the algebra in terms of $3\times 3$ upper triangular matrices
\be\label{basis31}
X_1=\begin{pmatrix} 0 &0&1 \\ 0&0&0 \\0&0&0 \end{pmatrix}, \quad 
X_2=\begin{pmatrix} 0 &1&0 \\ 0&0&0 \\0&0&0 \end{pmatrix}, \quad 
X_3=\begin{pmatrix} 0 &0&0 \\ 0&0&1 \\0&0&0 \end{pmatrix}.
 \ee
Then, any element of the corresponding group $A_{3,1}$ may be parametrized as 
\be \label{groupelement}
g=\begin{pmatrix} 1 &x^2&x^1 \\ 0&1&x^3 \\0&0&1 \end{pmatrix}.
\ee
This is clearly a non-compact group. According to the above discussion, in order to produce a compact manifold 
out of it, a compact discrete subgroup $\G$ has to be considered. Such a subgroup is given by those elements $g\in A_{3,1}$ which have integer values of $x^a$.
 Then the 
quotient $A_{3,1}/\G$ is indeed a compact nilmanifold. 

The Lie algebra invariant 1-form $e$ is given by
\be \label{e31}
e=\begin{pmatrix} 0&dx^2&dx^1-x^2dx^3 \\ 0&0&dx^3 \\ 0&0&0 \end{pmatrix}. \ee
Clearly, its components are 
\be\label{ea31} e^1=dx^1-x^2dx^3, \quad e^2=dx^2, \quad e^3=dx^3, 
\ee
which evidently satisfy the Maurer-Cartan equations, since $de^2=de^3=0$ and $de^1=-e^2\wedge e^3$.
The twist matrix has the form
\be\label{u31}
U=\begin{pmatrix}
1 & 0 & -x^2 \\
0 & 1 & 0 \\
0 & 0 & 1
  \end{pmatrix}
\ee 
and therefore the required identifications are
\be\label{id1}
 (x^1,x^2,x^3)\sim (x^1+1,x^2,x^3)\sim(x^1,x^2,x^3+1)\sim(x^1+x^3,x^2+1,x^3).
\ee 
From now on we shall refer to (\ref{id1}) (and similar ones) as twisted identifications. 

Moreover, it is straightforward to determine the vector fields 
$\tilde e_a$ which are dual to the 1-forms (\ref{ea31}).
They are 
\be
\tilde e_1 = \partial_1, \quad \tilde e_2=\partial_2, \quad \tilde e_3 = \partial_3+x^2\partial_1.
\ee
These do not deserve to be collectively called Killing vector fields since only $\tilde e_1$ generates 
an isometry, while the other two do not \cite{kstt}. However, it should be noted that these vector 
fields do not commute but instead they satisfy 
\be \label{vfcr}
[\tilde e_2, \tilde e_3]=\tilde e_1.
\ee
This observation will be important in section 3.

 Another simple observation is that rescaling the central element $X_1$ of the algebra by an integer, i.e. 
$X_1\rightarrow \frac{1}{N}X_1,~ N\in \Z$, leads to the commutation relation $$[X_2,X_3]=NX_1.$$
Then the effect on the above geometric data is that one has to replace $x^2$ in (\ref{e31})-(\ref{u31}) 
by $Nx^2$, while the last identification in (\ref{id1}) becomes $(x^1+Nx^3,x^2+1,x^3)$. This is related 
to the presence of quantized flux as we discuss below.

Let us briefly recall that the above twisted 3-torus can be also obtained by T-dualizing a square torus 
with $N$ units of NS 3-form flux, proportional to its volume form, turned on. Indeed, let us choose a gauge where 
the B
field is 
\be \label{b13}
B_{31}=Nx^2,
\ee 
and the metric is the standard metric on the 3-torus, $ds^2=\d_{ab}dx^adx^b$. Then one can use the Buscher rules \cite{Buscher:1987qj}
to perform a T-duality along the $x^1$ direction. For completeness, let us remind these rules for the 
metric and the B field when a T-duality is performed along the direction $i$~:
\bea
& G_{ii}\xrightarrow{T_i} \frac 1{G_{ii}}, \quad G_{ai} \xrightarrow{T_i} \frac{B_{ai}}{G_{ii}}, \quad 
G_{ab} \xrightarrow{T_i} G_{ab}-\frac{G_{ai}G_{bi}-B_{ai}B_{bi}}{G_{ii}}, \nn\\
& B_{ai}\xrightarrow{T_i} \frac{G_{ai}}{G_{ii}}, \quad B_{ab}\xrightarrow{T_i}
 B_{ab}-\frac{B_{ai}G_{bi}-G_{ai}B_{bi}}{G_{ii}},\label{br}
\eea
where $T_i$ denotes the T-duality action.
In the present case the result is that in the T-dual frame the B field 
vanishes and the dual metric corresponds exactly to that of the twisted 3-torus, i.e. it is given by 
$ds^2=\d_{ab}e^ae^b$ with the 1-forms as in (\ref{ea31}). Therefore, 
a twisted torus background is T-dual to a square torus background with non-constant B field. 
We shall return to this point again, after we will have studied matrix compactifications.
 
\subsection{A 6-dimensional example}

Let us now move on to a less simple 6-dimensional example. In six dimensions there are several nilpotent 
Lie algebras and therefore several cases of nilmanifolds. In fact, excluding algebras which are 
algebraic sums of lower-dimensional ones, there are 22 nilpotent Lie algebras{\footnote{Including 
the algebraic sums this number rises to 34 \cite{Grana}.}} up to isomorphism \cite{Patera:1976ud}. 

In the present subsection we consider the algebra $\mathcal A_{6,5}^{\a}$, where the superscript $\a$ 
denotes that there is an additional parameter in this case{\footnote{A different case was examined in 
\cite{Chatzistavrakidislie}.}. This algebra has the following commutation relations 
\be \label{cr65}
[X_1,X_3]=X_5, \quad [X_1,X_4]=X_6, \quad [X_2,X_3]=\a X_6, \quad [X_2,X_4]=X_5.
\ee 
A basis is given by the following $6\times 6$ 
upper triangular matrices:
\bea\label{basis63}
X_1&=&\begin{pmatrix} 0&0&0&1&0&0 \\ &0&0&0&0&0 \\ &&0&0&0&0 \\ &&&0&0&0 \\ &&&&0&0 \\&&&&&0 \end{pmatrix},
\quad 
X_2=\begin{pmatrix} 0&0&1&0&0&0 \\ &0&0&0&0&0 \\ &&0&0&0&0 \\ &&&0&0&0 \\ &&&&0&0 \\&&&&&0 \end{pmatrix},
 \nn\\
X_3&=&\begin{pmatrix} 0&0&0&0&0&0 \\ &0&0&0&0&0 \\ &&0&0&0&\a \\ &&&0&1&0 \\ &&&&0&0 \\&&&&&0 \end{pmatrix},
\quad
X_4=\begin{pmatrix} 0&0&0&0&0&0 \\ &0&0&0&0&0 \\ &&0&0&1&0 \\ &&&0&0&1 \\ &&&&0&0 \\&&&&&0 \end{pmatrix},
\nn \\ 
X_5&=&\begin{pmatrix} 0&0&0&0&1&0 \\ &0&0&0&0&0 \\ &&0&0&0&0 \\ &&&0&0&0 \\ &&&&0&0 \\&&&&&0 \end{pmatrix},
\quad 
X_6=\begin{pmatrix} 0&0&0&0&0&1 \\ &0&0&0&0&0 \\ &&0&0&0&0 \\ &&&0&0&0 \\ &&&&0&0 \\&&&&&0 \end{pmatrix}.\nn 
\eea
The general group element is found to be
\be
g=\begin{pmatrix} 1&0&x^2&x^1&x^5&x^6 \\ &1&0&0&0&0 \\ &&1&0&x^4&\a x^3 \\ &&&1&x^3&x^4 \\ &&&&1&0 \\&&&&&1 \end{pmatrix},
\ee
while the 1-form $e$ can be computed and it has the following form,
\be
e=\begin{pmatrix} 0&0&dx^2&dx^1&dx^5-x^2dx^4-x^1dx^3&dx^6-\a x^2dx^3-x^1dx^4 \\ &0&0&0&0&0 
\\ &&0&0&dx^4&\a dx^3 \\ &&&0&dx^3&dx^4 \\ &&&&0&0 \\&&&&&0 \end{pmatrix},
\ee
with components
\bea\label{ea65}
e^i&=&dx^i, \quad i=1,\dots,4, \nn\\  e^5&=&dx^5-x^2dx^4-x^1dx^3, \nn\\ e^6&=&dx^6-\a x^2dx^3-x^1dx^4.
\eea
Then the twist matrix is
\be 
U=\begin{pmatrix}
	1&0&0&0&0&0 \\
	0&1&0&0&0&0 \\
	0&0&1&0&0&0 \\
	0&0&0&1&0&0 \\
	0&0&-x^1&-x^2&1&0 \\
	0&0&-\a x^2&-x^1&0&1
  \end{pmatrix},
\ee
and the twisted identifications are found to be
\bea\label{id3} (x^1,x^2,x^3,x^4,x^5,x^6)&\sim&(x^1,x^2,x^3+c,x^4,x^5,x^6) \nn \\
&\sim&(x^1,x^2,x^3,x^4+d,x^5,x^6) \nn\\ &\sim&(x^1,x^2,x^3,x^4,x^5+e,x^6) \nn \\ 
&\sim&(x^1,x^2,x^3,x^4,x^5,x^6+f) \nn\\
&\sim&(x^1+a,x^2,x^3,x^4,x^5+ax^3,x^6+ax^4)\nn\\
&\sim&(x^1,x^2+b,x^3,x^4,x^5+bx^4,x^6+\a bx^3), \quad a,b,c,d,e,f\in\Z, \nn\\ \eea  
Under (\ref{id3}) we obtain the desired twisted compactification. 

The vector fields which are dual to the 1-forms (\ref{ea65}) in the present case read as
\bea 
\tilde e_i&=&\partial_i, \quad i=1,2,5,6, \nn\\
\tilde e_3&=&\partial_3+x^1\partial_5+\a x^2\partial_6, \nn\\
\tilde e_4&=&\partial_4+x^2\partial_5+x^1\partial_6.\label{vf65}
\eea
Of them, only $\tilde e_5$ and $\tilde e_6$ are Killing vector fields, i.e. they generate an isometry. 
The full set of vector fields (\ref{vf65}) satisfies the algebra (\ref{cr65}).

As before, the central elements $X_5$ and $X_6$ may be rescaled by integer numbers as 
$X_5\rightarrow \frac{1}{M}X_5$ and $X_6\rightarrow \frac{1}{N}X_6$, leading to the modified commutation 
relations 
\be 
[X_1,X_3]=MX_5, \quad [X_1,X_4]=NX_6, \quad [X_2,X_3]=\a NX_6, \quad [X_2,X_4]=MX_5.
\ee 
Then in (\ref{ea65}) $e^5$ and $e^6$ change to
\be \label{ea65b}
e^5=dx^5-Mx^2dx^4-Mx^1dx^3, \quad e^6=dx^6-\a Nx^2dx^3-Nx^1dx^4,
\ee
which modify the twist matrix and the twist identifications accordingly. 

The above background can also be obtained by T-dualizing a square 6-torus with appropriate quantized 
3-form fluxes. In particular, consider a 6-torus with the standard square metric, endowed with 
fluxes generated by the non-constant B field with values
\be
B_{53}=Mx^1, \quad B_{54}=Mx^2, \quad B_{63}=\alpha Nx^2, \quad B_{64}=Nx^1.
\ee 
The corresponding fluxes are 
\be\label{hflux65}
H_{153}=H_{254}=M \quad \mbox{and} \quad H_{263}=\a N, \quad  H_{164}=N.
\ee
Then one can use the Buscher rules (\ref{br}) to show that performing two consequtive T-dualities along the directions 
$x^5$ and $x^6$ the 6-dimensional nilmanifold described before is obtained, i.e. the B field vanishes 
and the metric is given by $ds^2=\d_{ab}e^ae^b$ with 1-forms as in (\ref{ea65}) and (\ref{ea65b}) respectively. 

The same procedure may be followed for any other nilmanifold in any dimension.
In the appendix A we collect the resulting 
twist matrices and twisted identifications for a class of nilmanifolds in five and six dimensions.

\section{Matrix theory compactifications}

\subsection{The BFSS and IKKT matrix models}

Let us start by briefly describing the two basic string-inspired matrix models (MMs), widely known as 
BFSS \cite{Banks:1996vh} and IKKT \cite{Ishibashi:1996xs}. 
The BFSS MM, also referred to as Matrix theory, was suggested as a non-perturbative definition of M theory. Its action, determining the dynamics of $N$ D0 branes in uncompactified 
space-time, is given by the following functional:
\be\label{BFSSaction}
{\cal S}_{BFSS}=\frac 1{2g}\int dt\biggl[ Tr\big(\dot  {\mathcal X}_a\dot {\mc X}_a
-\frac 12 [{\mc X}_a,{\mc X}_b]^2\big)
+2\psi^T\dot\psi-2\psi^T\Gamma^a[\psi,{\mc X}_a]\biggl], 
\ee  
where $ \mc X_a(t),a=1,\dots,9$ are nine time-dependent $N\times N$ Hermitian matrices, $\psi$ are their fermionic superpartners 
and $\G^a$ furnish a representation of $SO(9)$. In the following we shall be concerned mainly with the 
bosonic part of the above action.

The equations of motion resulting from the variation of the action (\ref{BFSSaction}) with respect to 
$ \mc X_a$, setting $\psi=0$, are 
\be \label{BFSSeom}
\ddot {\mc X}_a+[{\mc X}_b,[{\mc X}^b,{\mc X}_a]]=0,
\ee
where indices are raised and lowered with $\delta_{ab}$ and therefore it does not make any difference 
whether they are upper or lower. For static configurations it is clear that the first term in (\ref{BFSSeom}) 
may be dropped.

On the other hand, the IKKT MM was suggested as a non-perturbative definition of type IIB superstring 
theory and may be regarded as the D-instanton analog of the previous model. It is described by the 
following action:
\be \label{IKKTaction}
S  =  {\frac{1} {2g}}Tr\biggl(-{\frac 1 2}[{\mc X}_a,{\mc X}_b]^2
-\bar{\psi}\Gamma ^a[{\mc X}_a,\psi ]\biggl),
\ee
where now the number of matrices $\mc X_a$ is ten, namely the index $a$ takes the values $0,\dots,9$ and $\psi$ 
is a spinor of $SO(10)$  in the Euclidean  model where the indices are raised and lowered with the metric $\delta_{ab}$ .

The corresponding equations of motion are 
\be\label{IKKTeom} [{\mc X}_b,[{\mc X}^b,{\mc X}_a]]=0, \ee 
which are formally the same as the time-independent equations of the BFSS model{\footnote{Note however 
that in the IKKT model Euclidean time is also treated as a matrix variable.}}.

\subsection{Compactification on tori}

 In the following we shall consider compactifications 
of the above MMs, focusing on the BFSS model. Let us start by reviewing the cases of compactification on 
multidimensional tori and non-commutative versions of them \cite{cds}. Moreover, we supplement this discussion 
by providing the matrix analog of a square 3-torus with NS 3-form flux in the same approach.  This will provide the guidelines 
for the investigation of compactifications on twisted tori and their non-commutative versions, which
follows in the next subsections.

A matrix compactification on a $d$-dimensional torus is defined by 
a restriction of the matrix action under certain periodicity conditions incorporating the cycles of the 
torus. The simplest example involves compactification on a circle, but it is more illuminating for our 
purposes to start with the  case of a 3-torus.

For a T$^3$ extending, say, in the directions $\mc X_1,\mc X_2,\mc X_3$, the compactification involves three invertible 
unitary matrices $U_i$  obeying 
\bea \label{conditionst3}
\mc X_1+R_1&=&U_1\mc X_1U_1^{-1}, \nn\\
\mc X_2+R_2&=&U_2\mc X_2U_2^{-1}, \nn\\
\mc X_3+R_3&=&U_3\mc X_3U_3^{-1}, \nn\\
\mc X_a&=&U_i\mc X_aU_i^{-1}, \quad a\ne i, \quad a=1,\dots,9, \quad i=1,2,3,
\eea where $R_i$  are complex constants.

\paragraph{Compactification on T$^3$.} 

A simple solution of the conditions (\ref{conditionst3}) is given by 
\bea \label{sol1}
\mc X_i&=&iR_i\mathcal D_i,  
\quad \mc X_m=\mathcal A_m(U_i), \quad m=4,\dots,9, \nn\\
U_1&=&e^{ix^1}, \quad U_2=e^{ix^2}, \quad U_3=e^{ix^3}, \label{sln1}
\eea
where $x^i$ are coordinates on T$^3$ and $\mathcal D_i$ are covariant derivatives
\be 
\mathcal D_i=\partial_i-i\mathcal A_i(U_j),
\ee
with $\partial_i\equiv \partial/\partial x^i$.
This is in fact the unique solution, up to gauge equivalence, for the standard (i.e. commutative) 
3-torus. Note that in the 
present case it holds that 
\be\label{ucr3} [U_i,U_j]=0, \ee 
which is enough to guarantee that the $U$-dependence of the gauge potentials $\mathcal A$ ensures that the 
conditions (\ref{conditionst3}) are indeed satisfied.

Furthermore, substituting the above form of the solution in the bosonic sector of the BFSS action functional 
(\ref{BFSSaction}) 
one gets
\be \label{sym1}
\mathcal S=\frac 1{2g}\int dt\int d^3x Tr\biggl(-R_i^2\dot{\mc D}_i\dot{\mc D}_i+\dot{\mc A}_m\dot{\mc A}_m
+\frac 12 R_iR_j[\mc D_i,\mc D_j]^2-iR_i[\mc D_i,\mc A_m]^2-\frac 12[\mc A_m,\mc A_n]^2\biggl),
\ee
corresponding to the bosonic part of the action of a (1+3)-dimensional supersymmetric Yang-Mills (SYM) theory.

The above are readily generalized in $d$ dimensions. In that case one has to determine $d$ unitary matrices 
$U_i,i=1,\dots,d$ such that 
\bea \label{conditionstd}
\mc X_i+R_i&=&U_i\mc X_iU_i^{-1}, \nn\\
\mc X_a&=& U_i\mc X_aU_i^{-1}, \quad a\ne i, \quad a=1,\dots, 9, \quad i=1,\dots,d,
\eea 
with $d$ complex constants $R_i$. The solution is readily given by the direct generalization of the previous 
one, namely
\bea 
\mc X_i&=&iR_i\mathcal D_i,  \quad \mc X_m=\mathcal A_m(U_i), \quad m=d+1,\dots,9, \nn\\
U_i&=&e^{ix^i}, 
\eea
where $x^i$ are the coordinates of the torus T$^d$, while again the $U$s are commuting, i.e. (\ref{ucr3}) 
remains true.
 The corresponding $(1+d)$-dimensional SYM action is the direct generalization of (\ref{sym1}).

\paragraph{Compactification on non-commutative T$^3_{\theta}$.}

The conditions (\ref{conditionst3}) and (\ref{conditionstd}), apart from the solution on a standard 
3-torus and $d$-torus respectively, allow in fact for more general configurations \cite{cds}. Focusing on the 
general $d$-dimensional case, this may be easily 
seen by considering the operator
\be 
Q_{(ij)}=U_iU_jU_i^{-1}U_j^{-1}, \quad i\ne j.
\ee 
It is readily verified that, rather generally, this operator satisfies
\be\label{qxcr}
[Q_{(ij)},\mc X_a]=0, \quad \forall ~a=1,\dots,9.
\ee
Let us mention that the latter condition carries less information than (\ref{conditionstd}), since its 
derivation assumes an underlying associative structure. This will turn out to be important in certain 
instances below. 
Keeping this remark in mind,  from (\ref{qxcr}) it follows that $Q_{(ij)}$ is a scalar operator, which implies 
that
\be \label{ucr}
U_iU_j=\lambda_{ij}U_jU_i, \ee
with complex constants  $\lambda_{ij}=e^{2\pi i\theta^{ij}}$. The special case of $\theta^{ij}=0$, or 
equivalently $\lambda_{ij}=1$, corresponds to commuting $U$s and thus it implements the previously discussed 
case of the standard $d$-torus T$^d$. 

However, in general the $\theta^{ij}$ are not vanishing, in which case the $U$s are not commuting operators.
It is straightforward to verify \cite{Brace:1998ku} that a solution of the conditions (\ref{conditionstd}) in the present case 
is obtained for  
\bea \label{slnnctd}
\mc X_i&=&iR_i\hat{\mathcal D}_i,  \quad\mc X_m=\mathcal A_m(\hat U_i), \quad m=d+1,\dots,9, \nn\\
U_i&=&e^{i\hat x^i}
\eea
The following important remarks are in order. First of all, we have introduced hatted quantities in 
(\ref{slnnctd}) due to the fact that the gauge potentials cannot depend anymore just on $U$s as in the 
commutative case. In order for the conditions (\ref{conditionstd}) to be satisfied the dependence 
has to be on the set of operators $\hat U_i$ which commute with all $U_i$, i.e.
\be \label{cruu}
[\hat U_i, U_j] =0.
\ee
Moreover, a direct implication of (\ref{ucr}) is that the $\hat x^i$ do not commute as well, but instead 
they satisfy the relation
\be \label{3nc}
[\hat x^i, \hat x^j]=-2\pi i\theta^{ij}.
\ee
Thus they may be interpreted as the ``coordinates'' of a non-commutative $d$-torus T$^d_{\theta}$ 
(more precisely they should be called coordinate operators, since they are not coordinates in the classical 
sense).
Likewise, the covariant derivative $\hat{\mathcal D}_i$ is now defined as
\be \label{covder}
\hat{\mathcal D}_i=\hat\partial_i-i\mathcal A_i(\hat U_j).
\ee
Let us note that the operators $\hat\partial_i$ commute among themselves and therefore 
the vacuum solution for $\mc X_i$ satisfies the equations of motion (3.2), since $[\mc X_i, \mc X_j]=0$. This statement remains valid for all the solutions that we describe 
here and in the following.

The general procedure one may follow in order to determine the set of operators $\hat U_i$ was described in 
\cite{cds}. This includes cases where twisted gauge bundles are considered, which is beyond the 
scope of the present paper\footnote{In the richer case of twisted gauge bundles, the vacuum solution for $\mc X_i$ 
still satisfies the equations of motion (3.2), although in a less trivial way since $[\mc X_i, \mc X_j] \ne 0$.}. 
In the present case the $\hat U$s have the simple form
\be 
\hat U_i=e^{i\hat x^i-2\pi\theta^{ij}\hat\partial_j},
\ee 
and they satisfy the requirement (\ref{cruu}) as well as
\be 
\hat U_i\hat U_j=e^{2\pi i \hat\theta^{ij}}\hat U_j\hat U_i, \quad \hat\theta^{ij}=-\theta^{ij}. \ee
Therefore, the compactification on a non-commutative torus leads to a SYM theory where the gauge fields 
and scalars
$\mathcal A_a, a=1,\dots,9$ are fields on a dual non-commutative torus with parameter $\hat\theta$. 
Clearly, the same holds for fermions when they are included in the action. The form of this SYM action 
is similar to (\ref{sym1}), but with all commutators exchanged with $\star$-commutators, e.g. 
\be \label{stargauge}
[\mc A_m,\mc A_n] \longrightarrow [\mc A_m,\mc A_n]_{\star} \equiv \mc A_m\star\mc A_n-\mc A_n\star \mc A_m. 
\ee 
The $\star$ product is a deformation of the ordinary product for functions and it encodes the non-commutative algebraic information of the theory. In order to represent the (non-commutative) algebra  on the space of commuting coordinates, one needs to map the basis in the algebra\footnote{An algebra needs to  fulfil the
Poincar\' e-Birkoff-Witt property \cite{Madore:2000en}.} to the basis of monomials of commuting coordinates
. This map provides a representation of the elements of the abstract algebra in terms of functions of commuting coordinates, and these functions are then multiplied with the $\star$ product. Furthermore, one needs to map derivatives from the abstract algebra to the space of commuting coordinates.  

 In the present case of constant non-commutativity 
$\hat\theta^{ij}$ the $\star$ product is given by the well-known expression
\be 
f\star g = e^{\frac i2 \frac{\partial}{\partial x^i}2\pi\hat\theta^{ij}\frac{\partial}{\partial y^j}}f(x)g(y)|_{y\rightarrow x},
\ee 
corresponding to the Moyal-Weyl product, while the derivatives in the  algebra are mapped to the usual derivatives on the space of commuting coordinates.   

In the spirit of non-commutative geometry \cite{Connes:1994yd,Madore:2000aq}, the $U_i$ are the generators of an algebra 
$\mc A_{\theta}$ serving as the defining algebraic structure of the non-commutative torus. In analogy 
$\hat U_i$ comprise the algebra $\mc A_{\hat\theta}$ of the dual non-commutative torus, where the gauge 
theory resides. Moreover, $\hat {\mc D}$ serves as a linear connection defined on a projective module 
over $\mc A_{\hat\theta}$, the analog of a connection on a vector bundle in ordinary smooth geometry. 
We shall not go any further on these matters. Detailed expositions may be found for example in 
\cite{Connes:1994yd,cds}. 

\paragraph{Compactification on non-commutative T$^3_{\text{x}}$ .}

Apart from the solutions that we already reviewed, here we suggest that there are alternative ways to 
solve the conditions (\ref{conditionst3}) of the compactification on a 3-torus. These correspond to 
non-commutative tori with non-constant non-commutativity, which we denote as T$^3_{\text{x}}$.

Let us consider again the set of matrices (\ref{slnnctd}), specializing to the case of $d=3$, although 
our discussion may be directly generalized for any $d$.
Moreover, let us require that the only non-vanishing commutation relation of $\hat x^i$ is 
\be \label{slnnat3}
{[}\hat x^1, \hat x^3] =iN\hat x^2, \\
\ee
for an arbitrary scale $N$, while keeping the Heisenberg commutation relations $[\hat x^i, \hat\partial_j] =
-\d_j^i$. It is directly verified that the conditions (\ref{conditionst3}) are satisfied and therefore the 
above solution serves as a consistent background for the compactification of the matrix model. However, it is 
clear that (\ref{qxcr}) is not satisfied. This is due to the following subtlety. In the present case the 
associator 
$[\hat\partial_2,\hat x^1,\hat x^3]\equiv [\hat\partial_2,[\hat x^1,\hat x^3]]+\text{(cyclic permutations)}$, 
does not vanish. Indeed, we find 
\be \label{na}
[\hat\partial_2,\hat x^1,\hat x^3]=iN.
\ee
Due to this non-associativity, the relation (\ref{qxcr}) does not directly follow from the conditions 
(\ref{conditionst3}) anymore. Therefore, although the latter are satisfied, as they should, the former 
does not. This of course does not spoil the consistency of the solution.
Let it be stressed that the associator (\ref{na}) is the only non-vanishing one, while all the rest of the Jacobi identities still hold, e.g. $[x_i,x_j,x_k]=0, [\partial_i,\partial_j,\partial_k]=0, [\partial_i,\partial_j,x_k]=0$. This ensures the compatibility of the solution with the gauge invariance of the matrix model action. Indeed, the latter is based only on the  Jacobi identities $[U_i,[X_j,U^{-1}_i]]+\text{(cyclic permutations)}=0, [U_i,[X_j,X_k]]+\text{(cyclic permutations)}=0$  which stem from the vanishing associators and therefore they are satisfied.

According to the above, it is reasonable to interpret the resulting compactification as one on a 
non-commutative torus T$^3_{\text{x}}$ with an additional non-associative structure on the 
corresponding phase space. In the present case the non-commutativity is 
non-constant and the non-associativity a constant one. In section 4 we shall associate this solution 
to the background of a square 3-torus carrying an constant NS 3-form flux{\footnote{A different approach on this 
matter may be found in \cite{Anazawa:1999ub}, where instead the conditions (\ref{conditionst3}) are modified.}} which was described 
in section 2. It is worth mentioning that indications of non-associativity in backgrounds carrying 
a non-vanishing 3-form flux were given recently in a different context 
\cite{Lust:2010iy,Blumenhagen:2010hj,Blumenhagen:2011ph}.

For the solution at hand it is straightforward to check that the $U$s satisfy the commutation relations
\bea \label{ncUs}
[U_1,U_2]&=&0, \quad [U_2,U_3]=0, \nn\\
U_1U_3&=&e^{-iN\hat x^2}U_3U_1.
\label{ucrt3x}\eea
A set of operators $\hat U_i$ commuting with the $U$s is given by 
\be
\hat U_1=e^{i\hat x^1+N\hat x^2\hat\partial_3}, \quad \hat U_2=e^{i\hat x^2}, \quad \hat U_3=e^{i\hat x^3-N\hat x^2\hat\partial_1},
\ee
and they satisfy the commutation relations
\bea 
[\hat U_1,\hat U_2]&=&0, \quad [\hat U_2,\hat U_3]=0, \nn\\
\hat U_1\hat U_3&=&e^{iN\hat x^2}\hat U_3\hat U_1,
\label{uhatnat3}\eea
which are dual to (\ref{ucrt3x}). These operators generate the algebra of the dual 
 torus and they 
provide the dependence of the gauge potentials\footnote{More exactly, the gauge 
potentials of (\ref{covder}) should be modified in the present case. In particular, ${\cal A}_2(\hat U)$ should be replaced by $ {\cal\hat A}_2={\cal A}_2(\hat U)+i{\cal A}_2(\hat U)\partial_1+i{\cal A}_2(\hat U)\partial_3$ in order to ensure that the space of gauge potentials is well-defined under gauge transformations $\hat U$. This modification leaves the compactification conditions satisfied, up to gauge transformations. A more detailed account on the resulting gauge theory will be given elsewhere.}.

 The non-commutativity of the resulting theory may be 
again encoded in the appropriate $\star$ product, which does not have the Moyal-Weyl form due to 
the $\hat x$-dependence in (\ref{uhatnat3}). Indeed, the relevant $\star$ product is now given by
\be \label{nastar}
f\star g = e^{\frac i2 Nx^2\left(\frac{\partial}{\partial y^1}\frac{\partial}
{\partial z^3}-\frac{\partial}{\partial y^3}\frac{\partial}
{\partial z^1}\right)}f(y)g(z)|_{y,z\rightarrow x}~.
\ee
Note  that this $\star$ product  is associative\footnote{Non-associative 
$\star$ products were studied in \cite{Cornalba:2001sm} where they were also associated to non-vanishing H flux backgrounds.}. However, the non-associativity (\ref{na}) is mapped to the usual phase space by mapping the derivatives in the algebra to the usual derivatives on the space of commuting coordinates:
\be 
[\partial_j,x^i]=\delta^i_j,\quad [\partial_2,[x^1,x^3]_{\star}]=iN\neq 0.\ee 
It is worth mentioning that a class of non-associative but commutative gauge theories was 
studied in \cite{Ramgoolam:2003cs}. It would be interesting to examine in detail how the non-associativity encountered in the present context  manifests itself  within the gauge theory over the (dual) non-commutative torus (\ref{uhatnat3}).

It is reasonable to ask whether the background (\ref{slnnctd}) may be further generalized along the same 
lines, i.e. with non-constant non-commutativity. First of all, the conditions (\ref{conditionst3}) impose 
the constraint that the commutation relations between coordinates and momenta remain unchanged, i.e. 
$[\hat x^i, \hat\partial_j] =-\d_j^i$. 
Furthermore, in the present paper we discuss trivial gauge bundles 
and therefore we assume that $[\hat\partial_i,\hat\partial_j]=0$. In order to be able to provide a solid 
argument on possible extensions of (\ref{slnnat3}) let us also require that the commutators between 
the $\hat x^i$ 
are linear in $\hat x^i$. A final constraint, imposed again by the conditions (\ref{conditionst3}), is that 
there should exist a set of operators, constructed from $\hat x^i$ and $\hat\partial_i$, which commute 
with $\hat x^i$. The most general algebra respecting the above constraints and requirements is
\bea 
[\hat x^1, \hat x^2]&=&
2\pi i\theta^{12}, \nn\\
{[}\hat x^1, \hat x^3] &=&iN\hat x^2+2\pi i\theta^{13}-2\pi N\theta^{12}\hat\partial_1+2\pi N\theta^{23}\hat\partial_3,\nn
\\
{[}\hat x^2,\hat x^3]&=&
2\pi i\theta^{23}. 
\label{slnnat3gen}
\eea
The solution (\ref{slnnat3}), which was 
described before, is a special case of (\ref{slnnat3gen}) with all $\theta$s vanishing and it will be useful in section 4.

These commutation relations lead to
\bea\label{tt3cr1app}
U_1U_3&=&\lambda_{13} U_3U_1, \quad \lambda_{13}=
e^{-iN\hat x^2-2\pi i \theta^{12}+2\pi N\theta^{12}\hat\partial_1-2\pi N\theta^{23}\hat\partial_3}, \\
U_1U_2&=&\lambda_{12}U_2U_1, \quad \lambda_{12}= e^{
-2\pi i \theta^{12}},
\label{tt3cr2app} \\
U_2U_3&=&\lambda_{23} U_3U_2, \quad \lambda_{23} =e^{
-2\pi i \theta^{23}}\label{tt3cr3app}.\eea
One can construct  the operators $\hat U_i$ which commute with $U$s as $\hat U_i=e^{i\hat y^i}$, where 
\bea \label{genyy}
\hat y^1&=&\hat x^1-2\pi i\theta^{12}\hat\partial_2-2\pi i\theta^{13}\hat\partial_3-iN\hat x^2\hat\partial_3
+2\pi N\theta^{12}\hat\partial_1\hat\partial_3-2\pi N\theta^{23}(\hat\partial_3)^2, \nn\\
\hat y^2&=&\hat x^2-2\pi i\theta^{23}\hat\partial_3+2\pi i\theta^{12}\hat\partial_1, \nn\\
\hat y^3&=&\hat x^3+2\pi i\theta^{23}\hat\partial_2+2\pi i\theta^{13}\hat\partial_1+iN\hat x^2\hat\partial_1
+2\pi N\theta^{23}\hat \partial_3\hat\partial_1-2\pi N\theta^{12}(\hat\partial_1)^2.
\eea 
These operators satisfy dual ($\lambda\to 1/\lambda$) relations to (\ref{tt3cr1app}-\ref{tt3cr3app}).
Relaxing the requirements that were posed above, it is in principle possible to reach more general 
algebraic structures. However, this task is beyond the scope of this paper.

Finally, it is natural to ask whether non-associativity could be avoided in the present approach. The answer is 
yes; indeed one can describe a solution where the $\hat x^i$ satisfy (\ref{slnnat3}), 
 moreover it holds that
 \be\label{xpartial1}
 {[}\hat\partial_2,\hat x^1]=iN \hat\partial_3,
 \ee
while the $U_1$ in (\ref{slnnctd}) is modified to $U_1=e^{i\hat x^1+N\hat x^2\hat\partial_3}$. 
Then one obtains 
a consistent solution of the conditions (\ref{conditionst3}), while ensuring that all the Jacobi identities 
for the algebra of $\hat x^i$ and $\hat \partial_i$ are satisfied. However, in that case the $U$s turn out to 
be commutative, i.e. they generate a commutative algebra of functions. Thus the gauge theory turns out to be 
a commutative one and the information from the non-trivial commutation relations of $\hat x^i$ is lost 
at the level of the action. This situation is not interesting for our purposes and the corresponding 
solutions will not be further discussed.

\subsection{Compactification on twisted 3-tori}

Having discussed the compactification of the MMs on multidimensional tori and their non-commutative 
counterparts, the stage is all set to move on to the study of twisted tori. We shall follow 
the same lines as before, starting with the special case of a standard (i.e. commutative) twisted torus and 
subsequently treating 
the case of a non-commutative one. In the present subsection we focus on the simplest case of the 
twisted 3-torus, while in the next one higher-dimensional twisted tori are discussed.

\paragraph{Compactification on twisted $\tilde {\text{T}}^3$.} 
The simplest example of a twisted torus arises for $d=3$, 
as we described in section 2. In that case, a (twisted) compactification is achieved by imposing 
and solving an appropriately extended set of constraints, incorporating the twisted identifications (\ref{id1}). These twisted constraints involve three unitary matrices $U_1,U_2,U_3$ and they are
\bea \label{conditionstt3}
\mc X_1+R_1&=&U_1\mc X_1U_1^{-1}, \nn\\
\mc X_2+R_2&=&U_2\mc X_2U_2^{-1}, \nn\\
\mc X_3+R_3&=&U_3\mc X_3U_3^{-1}, \nn\\ 
\mc X_1+R_2\mc X_3&=& U_2\mc X_1U_2^{-1}, \nn\\
\mc X_a&=&U_i\mc X_aU_i^{-1}, \quad a\ne i, \quad a=1,\dots,9, \quad i=1,2,3, \quad (a,i)\ne (1,2).
\eea
The latter constraints generalize the ones for the square torus appearing in (\ref{conditionst3}), thus 
incorporating the twist of the 3-dimensional nilmanifold $\tilde {\text{T}}^3$. Indeed, as in the torus case the constraints 
 (\ref{conditionst3}) reflect the defining relations (\ref{t3id}) of T$^3$, likewise the constraints 
(\ref{conditionstt3}) are tantamount to the defining relations (\ref{id1}) of the nilmanifold. 
Therefore, the restriction of the matrix action under (\ref{conditionstt3}) defines a compactification 
on the $\tilde {\text{T}}^3$ (see also \cite{ramgoolam}).

A solution of the above constraints is now given by 
\bea \label{sltt3}
\mc X_i&=&iR_i\mathcal D_i, 
 \quad \mc X_m=\mathcal A_m,
 \quad m=4,\dots,9, \nn\\
U_1&=&e^{ix^1}, \quad U_2=e^{ix^2-\frac{R_2R_3}{R_1}x^1\partial_3}, \quad U_3=e^{ix^3}.
\eea
Let us comment on the above solution. First of all for this solution the coordinates $x^i$ are ordinary 
commutative coordinates on the twisted 3-torus $\tilde {\textrm{T}}^3$. 
Unlike the coordinates, the matrices $U_i$ do not commute but instead they satisfy a single non-trivial 
commutation relation,
\be\label{tw}
U_2U_3=e^{-i\frac{R_2R_3}{R_1}x^1}U_3U_2.
\ee 
This relation is not associated to any non-commutative properties of the manifold. It just reflects 
in the present framework the non-abelian nature of the algebra of vector fields along the three directions 
of the 
twisted torus, see eq.~(\ref{vfcr}), and therefore it is totally expected. 
Regarding the gauge potentials, note that ${\cal A}_1$ in ${\cal X}_1$ has to be modified to $\hat{\cal A}_1={\cal A}_1+i\frac{R_2R_3}{R_1}{\cal A}_3\partial_2$ in order to satisfy the fourth equation in (\ref{conditionstt3}). The same modification has to be done in the non-commutative  case that follows.
Furthermore, since the compactification manifold is commutative, the gauge potentials $\mc A_i$ and 
the scalar fields $\mc A_m$ depend directly on the commuting coordinates $x^i$.

\paragraph{Compactification on non-commutative twisted $\tilde {\text{T}}^3_{\text{x}}$.}

In analogy to the square tori, it is also 
possible to consider a solution of the conditions (\ref{conditionstt3}) where all the $U_i$ have the same form, namely $U_i=e^{i\hat x^i}$, but with non-commuting coordinates $\hat x^i$. This was already 
noticed and elaborated in \cite{ramgoolam}. Setting 
\bea\label{nc}
&& [\hat x^2,\hat x^3]=iR\hat x^1, \quad [\hat x^1,\hat x^j]=0, \nn\\
&& [\hat\partial_1,\hat x^2]=iR\hat\partial_3,  \quad  [\hat\partial_i,\hat x^i]=1,
\eea
where $R=\frac{R_2R_3}{R_1}$,
the conditions (\ref{conditionstt3}) are satisfied and (\ref{tw}) is reproduced. 
It is worth noting that although the present case resembles (\ref{slnnat3}), it is in fact very different. 
Here, due to (\ref{nc}), associativity is guaranteed and the compactification is on a non-commutative 
twisted torus.

As far as the dependence of the fields of the gauge theory is concerned, in the present non-commutative case 
they depend on the set of operators $\hat U$s commuting with the $U$s. This set is given as follows:
\bea \label{sltt3hatnc}
&& \hat U_1=e^{i\hat x^1}, \quad \hat U_2=e^{i\hat x^2+R\hat x^1\hat\partial_3}, \quad \hat U_3=e^{i\hat x^3-R\hat x^1\hat\partial_2},
\eea
satisfying the relation 
\bea \label{sltt3hatcm}
&& \hat U_2\hat U_3=e^{iR\hat x^1}\hat U_3 \hat U_2,
 \eea
dual to (\ref{tw}). 
Once more one can define a $\star$ product of the form
\be 
f\star g = e^{\frac i2 Rx^1\left(\frac{\partial}{\partial y^2}\frac{\partial}
{\partial z^3}-\frac{\partial}{\partial y^3}\frac{\partial}
{\partial z^2}\right)}f(y)g(z)|_{y,z\rightarrow x}\;,
\ee
to represent the algebra relations (\ref{nc}) on the space of commuting coordinates. However, in the present case the derivatives in the algebra are not mapped in the usual derivatives $\hat\partial_i\to \partial_i^*\neq\partial_i$. One constructs $\partial_i^*$ derivatives by comparing $(\hat\partial\hat f)(\hat x)$ and $(\partial_i^*\star f)(x)$ expanded in the appropriate basis. This results in a perturbative (in the non-commutative parameter) expression for $\partial_i^*$, which can be generalized to a closed expression in some cases. 
In the present example we obtain 
$\partial_2^*=\partial_2$, $\partial_3^*=\partial_3$, $\partial_1^*=\partial_1+ \frac{iR}{2}\partial_2\partial_3 +  {\cal O}(R^2)$.

\paragraph{Compactification on non-commutative twisted 
$\tilde {\text{T}}^3_{\theta,\text{x}}$.}

Let us finally go one step further and follow a general approach for the treatment of the conditions 
(\ref{conditionstt3}), which will lead us to more general solutions than the ones we obtained above for the
 twisted 3-torus. 
This may be made 
clear by considering once more the operator
\be 
Q\equiv Q_{(ij)} = U_iU_jU_i^{-1}U_j^{-1}.\nn
\ee
It is straightforward to show that 
\be 
[Q_{(13)},\mc X_a]=[Q_{(12)},\mc X_a]=0,\;\forall ~a,
\ee
however,
\be \label{q23}
[Q_{(23)},\mc X_1]=-R_2R_3Q_{(23)}\ne 0.
\ee
Therefore, we encounter a mixed situation, where the two operators $Q_{(12)}$ and $Q_{(13)}$ commute 
with all the $\mc X_a$, but the remaining one does not. It follows that the former ones are scalar operators 
and thus
\bea\label{tt3cr1}
U_1U_3&=&\lambda_{13} U_3U_1, \quad \lambda_{13}= e^{-2\pi i \theta^{13}}, \\
U_1U_2&=&\lambda_{12}U_2U_1, \quad \lambda_{12}= e^{-2\pi i \theta^{12}},
\label{tt3cr2}\eea
while this is not true for $Q_{(23)}$. The commutation relation between $U_2$ and $U_3$ does not have the 
same form and it is in fact expected to be $\hat x$-dependent as previously. Let us note that for $\lambda_{12}=\lambda_{13}=1$ 
we recover either the case of the commutative twisted torus or the one on non-commutative twisted torus 
with purely non-constant non-commutativity, both discussed above. Therefore any solution in the 
present case should reduce either to the solution (\ref{sltt3}) or to (\ref{nc}) in the limit of $\theta^{12}=\theta^{13}=0$.

Without further ado, let us write down the solution of the conditions (\ref{conditionstt3}) for general $\theta^{12}$ and $\theta^{13}$,
\bea \label{slntt3b}
\mc X_i&=&iR_i\hat{\mathcal D}_i,  
\quad \mc X_m=\mathcal A_m,
 \quad m=4,\dots,9, \nn\\
U_i&=&e^{i\hat x^i},
\eea
where 
\bea 
[\hat x^1, \hat x^2]&=&2\pi i\theta^{12},\nn \\
{[}\hat x^1, \hat x^3] &=& 2\pi i\theta^{13},\nn \\
{[}\hat x^2,\hat x^3]&=&iR\hat x^1 + 2\pi i\theta^{23}.\nn\\
{[}\hat\partial_1,\hat x^2]&=&iR\hat\partial_3, \label{nctt3gen}
\eea
with $R\equiv \frac{R_2R_3}{R_1}$ as before.
In the present case {\textit{all}} the $U_i$ do not commute among themselves. 
In particular, along with (\ref{tt3cr1}) and (\ref{tt3cr2}) we obtain
\bea\label{tt3ub}
U_2U_3&=&e^{-2\pi i\theta^{23}-iR\hat x^1}U_3U_2.\eea

For the solution (\ref{slntt3b}-\ref{nctt3gen}) we can find  the set of $\hat U$s which give the 
connection on a trivial gauge bundle. They have the form $\hat U_i=e^{i\hat y^i}$, where
\bea \label{gentwyy}
\hat y^1&=&\hat x^1-2\pi i\theta^{13}\hat\partial_3-2\pi i \theta^{12}\hat\partial_2, \nn\\
\hat y^2&=&\hat x^2+2\pi i \theta^{12}\hat\partial_1-iR\hat x^1\hat\partial_3-2\pi
 i\theta^{23}\hat\partial_3-\pi R\theta^{13}(\hat\partial_3)^2, \nn\\
\hat y^3&=&\hat x^3+2\pi i \theta^{13}\hat\partial_1+i R\hat x^1\hat\partial_2+2\pi i\theta^{23}\hat\partial_2+
2\pi R\theta^{13}\hat\partial_2\hat\partial_3+\pi R\theta^{12}(\hat\partial_2)^2.
\eea
We observe that in the latter case, the commutation relations of the coordinates $\hat x^i$ involve both 
constant and non-constant parts.

Summarizing, we were able to find consistent solutions of (\ref{conditionstt3}), corresponding to 
compactifications of Matrix theory on the ordinary twisted 3-torus or on different 
versions of non-commutative twisted 
3-tori. Let us remind at this point that these solutions solve the equations of motion (3.2), since $[\mc X_i, \mc X_j]=0$ 
in the vacuum. This remains true in the 6-dimensional case which follows.

\subsection{Compactification on twisted 6-tori}

The approach of the previous subsection for the matrix compactification on twisted 3-tori may be directly 
generalized to twisted tori of any dimension, such as the ones described in section 2.3 and appendix A. 
Higher-dimensional twisted tori comprise richer structures than the 3-dimensional 
case, since they generically include more than one twist and therefore they can be associated to 
string / M theory compactifications with more fluxes. Indeed, in section 2.3 we saw that the twisted torus based 
on the 6-dimensional nilpotent group $A_{6,5}^{\a}$ is T-dual to a square 6-torus with NS fluxes given 
in (\ref{hflux65}). In the present section we study the compactification 
of Matrix theory on this example.

Let us therefore consider the case of the twisted 6-torus constructed by the algebra ${\cal A}_{6,5}^{\a}$. 
For simplicity we set the parameter $\alpha=1$, since it does not affect the generality of the discussion.
Guided by the 
identifications (\ref{id3}) it is easy to write down the necessary constraints, which now involve 
six unitary matrices $U_i,i=1,\dots,6$ and they take the form
\bea \label{conditionstt6}
\mc X_i+R_i&=&U_i\mc X_iU_i^{-1},  \nn\\
\mc X_5+R_1\mc X_3&=&U_1\mc X_5U_1^{-1}, \nn\\
\mc X_5+R_2\mc X_4&=&U_2\mc X_5U_2^{-1}, \nn\\
\mc X_6+R_1\mc X_4&=&U_1\mc X_6U_1^{-1}, \nn\\
\mc X_6+R_2\mc X_3&=&U_2\mc X_6U_2^{-1}, \nn\\
\mc X_a&=&U_i\mc X_aU_i^{-1},\quad a\ne i,  \quad (a,i)\ne (5,1), (5,2), (6,1), (6,2).
\eea
The solutions we consider involve again trivial gauge bundles and therefore we choose the $\mc X_a$ to be 
\be\mc X_i=iR_i\hat{\mc D}_i \quad \text{and} \quad \mc X_m=\mc A_m, \quad m=7,\dots,10,\label{slntt6}\ee
where the gauge potentials in the hatted covariant derivative generically depend on some $\hat U_i$ which commute 
with all the $U_i$, as before. More precisely, ${\cal A}_i$ are modified according to the relation $\hat {\cal A}_i={\cal A}_i+i\frac{R_jR_k}{R_i}f^{jk}_{\ \ i}{\cal A}_k\partial_j,j<k$.

\paragraph{Commutative twisted $\tilde {\text T}^6$.}

The commutative case is relatively easy to describe. It amounts to choosing the following set of unitary operators,
\bea
U_1&=&e^{ix^1-\frac{R_1R_3}{R_5}x^5\partial_3-\frac{R_1R_4}{R_6}x^6\partial_4}, \nn\\
U_2&=&e^{ix^2-\frac{R_2R_4}{R_5}x^5\partial_4-\frac{R_2R_3}{R_6}x^6\partial_3}, \nn\\ 
U_s&=&e^{ix^s}, \quad s=3,4,5,6.
\eea
Here $x^i$ are ordinary commutative coordinates and it is straightforward to show that the only non-trivial 
commutation relations for the $U_i$ are
\bea 
U_1U_3&=&e^{-i\frac{R_1R_3}{R_5}x^5}U_3U_1, \quad U_1U_4= e^{-i\frac{R_1R_4}{R_6}x^6}U_4U_1, \nn\\
U_2U_3&=&e^{-i\frac{R_2R_3}{R_6}x^6}U_3U_2, \quad U_2U_4=e^{-i\frac{R_2R_4}{R_5}x^5}U_4U_2.
\label{crutt6}\eea 
As we also discussed is the 3-dimensional case, for the commutative manifold these relations just reflect the 
non-abelian nature of the algebra of vector fields on it, which were described in (\ref{vf65}). 
Therefore, the compactification here is on an ordinary manifold, the gauge potentials and the scalar fields 
depend directly on the commutative coordinates $x^i$ and the resulting theory is a (1+6)-dimensional 
YM theory.

\paragraph{Non-commutative twisted $\tilde {\text T}_{\theta,\text{x}}^6$.}

After the brief exposition of the commutative case let us now turn our attention to the general case. 
Considering the operators $Q_{(ij)}$, one obtains that the ones which do not commute with $\mc X_a$ are 
the ones with the index pairing $(i,j)=\{(1,3),(1,4),(2,3),(2,4)\}\equiv \mc I$, which satisfy
\bea
[Q_{(13)},\mc X_5]&=&-R_1R_3Q_{(13)}, \quad [Q_{(14)},\mc X_6]=-R_1R_4Q_{(14)}, \\
{[}Q_{(23)},\mc X_6]&=&-R_2R_3Q_{(23)}, \quad [Q_{(24)},\mc X_5]=-R_2R_4Q_{(24)}.
\eea
The rest of the commutators among $Q_{(ij)}$ and $\mc X_a$ vanish. This means that 
\be 
U_iU_j=e^{2\pi i \theta^{ij}}U_jU_i, \quad (i,j)\notin \mc I.
\ee
Then we may consider the solution (\ref{slntt6}) with $U_i=e^{i\hat x^i}$ and the following commutation 
relations:
\bea
[\hat x^i,\hat x^j]&=&iR_{(ij)}f^{ij}_{\ \ k}\hat x^k+2\pi i\theta^{ij}, \nn\\
{[}\hat\partial_i,\hat\partial_j]&=&0, \nn\\
{[}\hat\partial_i,\hat x^j]&=&\d^{j}_i+iR_{(jk)}f^{jk}_{\ \ i}\hat\partial_k, \quad j<k~.\label{crtt6}
\eea
The $f_{ij}^{\ \ k}$ are the structure constants of the algebra $\mc A_{6,5}$, 
\be\label{sc65}
f_{13}^{  \ \ 5}=f_{14}^{\ \ 6}=f_{23}^{\ \ 6}=f_{24}^{\ \ 5}=1,
\ee
which are antisymmetric in the lower indices. Moreover, the quantities $R_{(ij)}$ in the example at hand are
\be
R_{(13)}=\frac{R_1R_3}{R_5}, \quad R_{(14)}=\frac{R_1R_4}{R_6}, \quad R_{(23)}=\frac{R_2R_3}{R_6}, 
\quad R_{(24)}=\frac{R_2R_4}{R_5}.
\ee
Clearly, the subscripts in parentheses are labels and not indices and therefore they are not summed in (\ref{crtt6}). Furthermore, it holds that $R_{(ij)}=R_{(ji)}$.

It is straightforward to check that the above set of $U_i$ and $\mc X_a$ furnishes a consistent solution 
of (\ref{conditionstt6}). The last relation in (\ref{crtt6}) is crucial for this consistency and moreover it 
guarantees the associativity of the full algebra of coordinates and momenta. However, apart from the 
above, it is important to be able to construct the set of operators $\hat U_i$ which commute with 
$U_i$. This is only possible upon imposing some additional constraints. Indeed, these operators can be 
written as $\hat U_i=e^{i\hat y^i}$, where 
\bea \label{ygen}
\hat y^i=\hat x^i-2\pi i\theta^{ij}\hat\partial_j-iR_{(ij)}f^{ij}_{\ \ k}\hat x^k\hat \partial_j
+\pi R_{(ij)}f^{i\{j}_{\ \ \ k}\theta^{l\}k}\hat\partial_l\hat\partial_j,
\eea
where the notation $\eta^{\{j}\xi^{l\}}$ for the superscripts denotes symmetrization or antisymmetrization 
as follows
\be
\eta^{\{j}\xi^{l\}}= \left\{
\begin{array}{ll}
\eta^{j}\xi^{l}-\eta^{l}\xi^{j}, & \mbox{for} \quad i=1,2, \quad (j,l)\in \mc I, \\
\eta^{j}\xi^{l}+\eta^{l}\xi^{j}, & \mbox{for} \quad i=3,4, \quad (j,l)\in \mc I, \\
\eta^j\xi^l, & \mbox{for} \quad (j,l)\notin \mc I
\end{array}
\right.\nn
\ee 
For $i=5,6$ the related term is absent due to (\ref{sc65}). The $\hat U_i$ constructed in that way commute 
with the $U_i$ under the following conditions:
\bea
\theta^{56}&=&0, \nn\\
\frac{\theta^{36}}{R_3R_6}&=&\frac{\theta^{45}}{R_4R_5}, \qquad 
\frac{\theta^{15}}{R_1R_5}=\frac{\theta^{26}}{R_2R_6}, \nn\\
\frac{\theta^{16}}{R_1R_6}&=&\frac{\theta^{25}}{R_2R_5}, \qquad 
\frac{\theta^{35}}{R_3R_5}=\frac{\theta^{46}}{R_4R_6}.
\eea
It is worth noting that an appropriate $\star$ product may be defined in the present case as well. 
It is of a mixed form and it is given by
\be 
f\star g = e^{\frac i2 \big(R_{(ij)}f^{ij}_{\ \ k}x^k+2\pi\theta^{ij}\big)\frac{\partial}{\partial y^i}\frac{\partial}
{\partial z^j}}f(y)g(z)|_{y,z\rightarrow x}~.
\ee
Due to the last relation in (\ref{crtt6})  the derivatives in the algebra are
not mapped to the usual derivatives, but instead $\hat\partial_i\to\partial_i^{*}=\partial_i+\frac{i}{2}R_{(jk)}f^{jk}_{\ \ i}\partial_j\partial_k+\mc O(R^2),~j<k$.
   
The above procedure may be directly generalized for the compactification of Matrix theory on any higher-dimensional 
nilmanifold, such as the 5- and 6-dimensional ones of appendix A. Then, the 
general expressions (\ref{crtt6}) and (\ref{ygen}) will still hold for an 
appropriate set of index pairs $\mc I$ and the structure constants  
 $f_{ij}^{\ \ k}$ of the corresponding nilpotent algebra.

\section{Dualities and Seiberg-Witten maps}

Having studied several backgrounds corresponding to compactifications of the BFSS model in section 3, let us 
now discuss how they may be related among themselves as well as with known backgrounds of M theory / type IIA 
string theory in the supergravity approximation. It is useful at this stage to summarize the solutions 
that 
we have described so far in order to facilitate their comparison. This is done in table 1.

\begin{table}[ht]
\begin{center}
 \begin{tabular}{l*{4}{c}c}
 & $[\hat x^i,\hat x^j]$      & $Q_{(ij)}$  & Torus type & Twist & SuGra Flux 
 \\
\hline \\
1&0  & 1 &  
3d C & No & 
-
\\ \\
2&$-2\pi i \theta^{ij}$         &  $e^{2\pi i\theta^{ij}}$ & 3d NC & 
No & $B_{ij}$  \\ \\
3&$iN\epsilon_{ij2}\hat x^2$          & $e^{-iN\epsilon_{ij2}\hat x^2}$ &  
3d NC,NA
& No & $H_{123}$\\ \\
4&0    & $e^{-i\frac{R_2R_3}{R_1}\epsilon_{ij1} x^1}$ & 3d C & 
Yes & $f^{\ \ 1}_{ 23}$ \\ \\
5&$i\frac{R_2R_3}{R_1}\epsilon_{ij1}\hat x^1$ & $e^{-i\frac{R_2R_3}{R_1}\epsilon_{ij1}\hat x^1}$ & 3d NC & Yes & $H_{123},f^{\ \ 1}_{ 23}$ \\ \\
6&$i\frac{R_2R_3}{R_1}\epsilon_{ij1}\hat x^1+2\pi i\theta^{ij}$	
&$e^{-i\frac{R_2R_3}{R_1}\epsilon_{ij1}\hat x^1-2\pi i\theta^{ij}}$	&3d NC	&	Yes&	$B_{ij},H_{123},f^{\ \ 1}_{ 23}$		\\ \\
7&0	&$e^{-iR_{(ij)}f^{ij}_{\ \ k}\hat x^k}$	&6d C	&	Yes&	$f_{ij}^{\ \ k}$	\\ \\
8&$iR_{(ij)}f^{ij}_{\ \ k}\hat x^k+2\pi i\theta^{ij}$	
&$e^{-iR_{(ij)}f^{ij}_{\ \ k}\hat x^k-2\pi i\theta^{ij}}$	&	6d NC&Yes	
& $B_{ij},H_{ijk},f_{ij}^{\ \ k}$
\end{tabular} 
\end{center}
\caption{Solutions of the BFSS model compactified on 3- and 6-dimensional tori and twisted tori.
 C stands for commutative, NC stands for non-commutative and NA for non-associative. Indices run 
from 1 to 3 for the 3-dimensional cases and from 1 to 6 for the 6-dimensional ones. The last 
column contains the associated supergravity fluxes for each compactification and it is discussed in section 4.}
\end{table}

\paragraph{Connes-Douglas-Schwarz conjecture.}

The starting point is that the BFSS model corresponds to a non-perturbative definition of M theory in the 
infinite momentum frame
\cite{Banks:1996vh}. Therefore one expects that known backgrounds of 11-dimensional supergravity (the 
field theory limit of M theory) should be reproduced in the matrix framework. Indeed, Connes, 
Douglas and Schwarz suggested that that the deformation parameters $\theta^{ij}$ defining the non-commutative 
tori as in section 3.2, correspond to moduli of the 11-dimensional supergravity \cite{cds}. The latter contains a 
3-form $C_{IJK}$, which is the gauge potential of the 4-form field strength of the theory, where $I,J,K$ are 
11-dimensional indices. Then the claim is that 
\be \label{cdsm}
\theta^{ij}\propto \int dx^idx^j C_{ij-},
\ee
where ``$-$'' denotes the light cone direction $x^-$ \cite{cds}. It is also useful to rephrase this 
statement in the language of the type IIA
theory, which is obtained by 11-dimensional supergravity upon compactification on a circle. In that 
process the 3-form $C$ gives rise to the NS 2-form field B of the type IIA supergravity{\footnote{We do not discuss here issues related to the Ramond-Ramond forms of the type IIA theory. A related discussion may be found in \cite{Brace:1998xz}.}}. 
Therefore, in IIA language, (\ref{cdsm}) may be restated as
\be \label{cds}
\theta^{ij}\propto \int dx^idx^j B_{ij}.
\ee 
In the following we shall retain this auxilliary type IIA language in our discussion.

According to the above, the deformation of a commutative torus to a non-commutative one in the matrix 
model corresponds to 
turning on background values for the B field in type IIA string theory. Let us now use this statement as 
a guiding principle in order to unveil relations between the backgrounds we studied in section 3 
and known type IIA backgrounds{\footnote{We use here the term ``background'' in a somewhat loose sense.
Some of the situations we discuss are not fully consistent string backgrounds and need to be 
appropriately lifted \cite{kstt}. However, here we are interested in relations between fluxes and deformations and 
this discussion in beyond our scope.}}. For this purpose we will use the following notation: when we want to refer 
to a solution of the compactified matrix model, such as the ones appearing in table 1, we write e.g.
``MM on $\text{T}_{\theta}^3$'' for the solution on the non-commutative 3-torus with constant non-commutativity 
$\theta^{ij}$. Similarly, for the auxiliary type IIA background we write e.g. ``IIA on $\tilde{\text{T}}^3$'' for 
a compactification on the twisted 3-torus.

In the above notation, the Connes-Douglas-Schwarz conjecture reads as
\be \label{cdssch}
\mbox{MM on}~ \mbox{T}_{\theta}^3 \quad \overset{CDS}\longleftrightarrow \quad \mbox{IIA on}~ \mbox{T}_{B}^3,
\ee 
corresponding to (\ref{cds}). The left hand side (lhs) of the conjecture refers to the solution 
(\ref{slnnctd}) described in section 3.2, which is solution 2 in table 1. This correspondence 
was further justified in \cite{Douglas:1997fm,Kawano:1998re}.

Along the same lines, we suggest the 
following correspondences 
for the solutions which appear in table 1:
\begin{itemize}
 \item The solution 
3 of table 1 describes the compactification of the BFSS model on a torus with 
non-constant non-commutativity. On the supergravity side this should correspond to a background with 
non-constant B field and therefore a constant 3-form flux H. Schematically,
\be\label{cdssch2}
\mbox{MM on}~ \mbox{T}_{x}^3 \quad \overset{CDS}\longleftrightarrow \quad \mbox{IIA on}~ \mbox{T}_{H}^3,
\ee
The rhs of this correspondence was already referred to in section 2.2, eq. (\ref{b13}). In other words 
we suggest that deforming the torus to a non-commutative one with non-constant non-commutativity 
corresponds to turning on a constant 3-form flux through the torus on the supergravity side. In 
11-dimensional language this situation corresponds to a constant 4-form background.
\item Turning to the compactification of the BFSS model on twisted 3-tori, the solution 
5 of table 1 should be associated to 
a non-constant B field on a twisted 3-torus on the supergravity side, 
\be\label{cdssch3}
\mbox{MM on}~ \tilde{\mbox{T}}_{x}^3 \quad \overset{CDS}\longleftrightarrow \quad \mbox{IIA on}~ 
\tilde{\mbox{T}}_{H}^3,
\ee 
This was discussed in  detail in \cite{ramgoolam}, while supergravity backgrounds with both geometric fluxes 
and NS fluxes were studied for example in \cite{km,Hull:2005hk,Hull:2006tp}. Moreover, in the present study we described more general solutions 
associated to twisted 3-tori, given by (\ref{slntt3b}) and (\ref{nctt3gen}). On the supergravity side, 
these should be associated to twisted 3-tori with mixed (constant and non-constant) B field.
\item Finally, as far as the compactification on a twisted 6-torus is concerned, the situation is very similar.
In particular, for the solution 8 in table 1 and vanishing $\theta^{ij}$,
\be\label{cdssch4}
\mbox{MM on}~ \tilde{\mbox{T}}_{x}^6 \quad \overset{CDS}\longleftrightarrow \quad \mbox{IIA on}~ 
\tilde{\mbox{T}}_{H}^6,
\ee 
where the NS fluxes H are given in (\ref{hflux65}). 
\end{itemize}

The above correspondences are plausible in view of previous work on the subject but in order to be 
fully demonstrated one has to study in detail the resulting (1+d)-dimensional theory.
We plan to perform such an analysis in a forthcoming 
publication.

\paragraph{T-duality and Seiberg-Witten maps.} 

In the previous paragraph we discussed relations of matrix model backgrounds with certain supergravity ones. 
However, here we suggest that there exist also relations among the matrix backgrounds themselves and among the gauge theories on them.

Looking at the solution of the constraints for the compactification on a non-commutative torus $\mathrm{T}_{\theta}^3$, eqs.~(\ref{slnnctd}) and (\ref{3nc}), we observe that there exists a mapping from the non-commutative torus to a commutative one:
\be \label{sw1}
f:\hat x^i\to x^i-\pi i\theta^{ij}\partial_j,\quad f:\hat\partial_i\to\partial_i,
\ee
where $\hat x^i$ and $\hat\partial_i$ are the coordinates and the corresponding derivatives  on 
the non-commutative torus, while $x^i$ and $\partial_i$ are the usual commuting coordinates and derivatives. 
Under this mapping the $U$s and $\hat U$s of the solution (\ref{slnnctd}) go to:
\be \label{newUs}
f:U_i=e^{i\hat x^i}\to e^{i x^i+\pi\theta^{ij}\partial_j},\quad
f:\hat U_i=e^{i\hat x^i-2\pi\theta^{ij}\hat\partial_j}\to e^{i x^i-\pi\theta^{ij}\partial_j},
\ee
preserving the algebras of $U$s and $\hat U$s. Thus the mapping (\ref{sw1}) induces the Seiberg-Witten (SW) map, i.e. the  
  transformation from  non-commutative Yang-Mills fields (\ref{stargauge}) to ordinary Yang-Mills fields (\ref{sol1}) over the same commutative torus. This construction of the SW map goes along the lines of the original construction introduced in the seminal paper \cite{SW}.  More abstractly, it is enough to assume that the non-commutative gauge transformation is induced by the ordinary one \cite{Madore:2000en}. One uses the representation of the elements of the non-commutative algebra in terms of functions of commuting
coordinates which are multiplied with the $\star$ product. Assuming that 
the non-commutative gauge transformation is induced by the ordinary one
provides enough data to express the non-commutative fields and gauge parameter as functions of the commutative ones. As the analysis of the Yang-Mills theory resulting from the compactification is beyond the scope of this paper we do not construct the SW map(s) explicitly.

A similar map can be constructed in the non-associative case, i.e. for the compactification on T$_{\rm x}^3$ defined by the relations (\ref{slnnat3})-(\ref{ncUs}).
There exists a mapping from the aforementioned solution of (\ref{conditionst3}) into  the  solution of the same condition but on a commutative torus:
\be \label{sw2}
f:\hat x^1\to x^1+\frac{iN}{2} x^2\partial_3,\quad f:\hat x^2\to x^2,\quad f:\hat x^3\to x^3-\frac{iN}{2} x^2\partial_1,\quad f:\hat\partial_i\to\partial_i,
\ee
where
\be 
[x^i,x^j]=0,\quad [\partial_i,\partial_j]=0,\quad [\partial_i,x^i]=1,\quad 
[\partial_2,x^3]=\frac{iN}{2} \partial_1,\quad [\partial_2,x^1]=-\frac{iN}{2} \partial_3.
\ee 
Note that the last two relations imply
$$ [\partial_2,x^1, x^3]=iN. $$
The operators $U_i$ and $\hat U_i$ are mapped to
\bea 
&& U_1=e^{i x^1-\frac{N}{2} x^2\partial_3},\quad  U_2=e^{i x^2},\quad U_3=e^{i x^3+\frac{N}{2} x^2\partial_1},\nn \\
&& \hat U_1=e^{i x^1+\frac{N}{2} x^2\partial_3},\quad \hat U_2=e^{i x^2},\quad \hat U_3=e^{i x^3-\frac{N}{2} x^2\partial_1},
\eea
and satisfy the same algebras as before. This induces  the SW map between gauge theories over non-commutative and commutative tori with the (same) non-associativity in the phase space.
It is easy, but not very illuminating, to construct a similar mapping  for the more general solution described by relations (\ref{slnnat3gen})-(\ref{genyy}).

As a final example of the relation between the gauge field theories over the compact spaces we discuss, we provide a map for a twisted compactification given by the relations
 (\ref{slntt3b})-(\ref{gentwyy}):
\be \label{sw3}
f:\hat x^i\to x^i+\pi i\theta^{ij}\partial_j+i\delta^{i3}Rx^1\partial_3,\quad f:\hat\partial_i\to\partial_i.
\ee
In the gauge sector this map induces the SW map between non-commutative and ordinary Yang-Mills theories over a (commutative) twisted torus.

Let us close this section with the following observation. It is well-known that there exists a T-duality among a square torus with non-constant background B field, 
i.e. with constant H flux, and a twisted torus with vanishing B field. This T-duality was briefly reviewed 
in section 2.2. Schematically, this means that 
\be\label{tdual}
\mbox{IIA on}~ {\mbox{T}}^3_{H} \quad \overset{T}\longleftrightarrow \quad \mbox{IIB on}~ 
\tilde{\mbox{T}}^3,
\ee 
where T-duality relates Type IIA and type IIB string theory as usual.
We observe that the lhs of the T-duality appears in (\ref{cdssch2}), while the rhs is directly associated to 
the solution (\ref{sltt3}) on a commutative twisted 3-torus (solution 4 in table 1). This allows us to construct the following 
diagram,
\[\renewcommand{\arraystretch}{2.0}
\begin{array}{ccccc}
\mbox{IIA on}~ {\mbox{T}}^3_{H} & \overset{T}\longleftrightarrow &\mbox{IIB on}~ 
\tilde{\mbox{T}}^3\\
\Biggl\updownarrow& & \Biggl\updownarrow
\\
\mbox{MM on}~ {\mbox{T}}_{x}^3&\longleftrightarrow&\mbox{MM on}~ \tilde{\mbox{T}}^3.
\end{array}\]
The vertical arrow on the lhs of this diagram is the correspondence (\ref{cdssch3}), while the one on the 
rhs simply relates two situations without fluxes or deformations. 
Then the horizontal arrow between the two MM solutions provides a
possible realization of T-duality at the level of Matrix theory, which deserves further 
investigation.
A similar diagram holds for the 6-dimensional case as well, where the T-duality is performed in two different 
directions as explained in section 2.3.

\section{Conclusions}

In the present paper we studied compactifications of Matrix theory on twisted tori and non-commutative versions of them. Our starting point was the construction of twisted tori realized as nilmanifolds based on  nilpotent Lie algebras.  Certain explicit examples were provided and their T-duality to square tori endowed with constant NS 3-form flux was discussed. Next, the toroidal compactification of the BFSS matrix model was revisited. Apart from the previously obtained results \cite{cds}, we described a solution of the compactification conditions which corresponds to a non-commutative deformation of the torus with non-constant non-commutativity. This solution carries an underlying non-associative structure on the corresponding phase space.
Thenceforth we moved on to study compactifications on twisted tori. Analysing the case of the twisted 3-torus, we identified
a set of solutions to the corresponding conditions  for commutative and non-commutative twisted 
3-tori. A similar analysis was carried out for a particular 6-dimensional twisted torus leading to solutions 
which were presented in a form allowing direct generalization to any other higher-dimensional nilmanifold.

In addition, we presented arguments relating known backgrounds of M theory / type IIA string theory in the supergravity approximation to the solutions of the BFSS model corresponding to the compactifications  we  
studied. In particular, along the lines of the Connes-Douglas-Schwarz correspondence \cite{cds}, 
non-commutative deformations of tori and twisted tori were associated to turning on fluxes in 11-dimensional 
supergravity. Moreover, star products associated with the corresponding non-commutative algebras were 
constructed and relations connecting non-commutative and commutative backgrounds, inducing the Seiberg-Witten 
map between the corresponding gauge theories, were determined. Finally, we indicated a possible realization of 
T-duality between twisted and untwisted tori in Matrix theory. However, these issues should be addressed in
 more detail by analysing the resulting gauge theories. In this process the spectra of BPS states 
should be carefully studied, along the lines of \cite{cds,Ganor:1996zk,ramgoolam}. We plan to report on this issue on a future publication.

An interesting future direction along the lines of the present paper would be to identify ways to describe 
analogs of non-geometric backgrounds in the framework of Matrix theory. Such backgrounds arise by performing 
a T-duality along the base of twisted tori instead of the fiber \cite{kstt}. Relations between non-geometry,
T-folds \cite{Hull:2004in} and 
non-commutativity were already reported in \cite{Lust:2010iy,Blumenhagen:2011ph,Grange:2006es}. 
Moreover, in recent work \cite{Andriot:2011uh} it was argued that a ten-dimensional perspective 
of non-geometric fluxes may be gained by describing backgrounds in terms of variables yielding the 
geometry globally well-defined. This path goes through generalized geometry and uses an antisymmetric 
bivector field as a sign of non-geometry. It would be interesting to investigate whether such a 
bivector can be traced in a non-commutative deformation of the compactification manifold in 
Matrix theory.

\vspace{10pt}

\paragraph{Acknowledgements.}

We would like to thank D. Andriot for useful discussions.
This work was partially supported by the SFB-Tansregio TR33
``The Dark Universe" (Deutsche Forschungsgemeinschaft),
the European Union 7th network program ``Unification in the
LHC era" (PITN-GA-2009-237920), the Ministry of
Science, Education and Sport of the Republic
of Croatia under the contract 098-0982930-2861, and the Alexander von Humboldt Foundation.

\appendix

\section{Geometric data for higher-dimensional nilmanifolds}

In this appendix we collect the twist matrices and the identifications for a class of nilmanifolds. 
These data are useful in order to fully determine the geometry of each case and classify the 
associated geometric fluxes. 

Before proceeding let us explain the requirements which single out the cases which we shall present. 
According to the tables of \cite{Patera:1976ud}, there exists a certain number of isomorphism classes of
 nilpotent Lie algebras in each dimension which are not algebraic 
sums of lower-dimensional ones.  We focus our attention on such cases. They include one 
3-dimensional case, which was treated in the main text, one 4-dimensional case, six 5-dimensional 
and 22 6-dimensional cases. Out of the latter, the $\mathcal A_{6,5}^{\a}$ was treated in the main text. 
Here we shall not present all the above cases. Instead we find it reasonable to impose the  
restriction that the Lie algebra satisfies the equations of motion of the BFSS and IKKT matrix models, 
in the former case at least for time-independent backgrounds,
\be 
\label{IKKTeom2} [X_b,[X^b,X_a]]=0 \quad \Leftrightarrow \quad f_c^{ab}f_b^{dc}X_d=0.
\ee
Such cases were studied from a different perspective in \cite{Chatzistavrakidislie}. It turns out that the 
relevant algebras are the $\mathcal A_{5,1}, \mathcal A_{5,4}, \mathcal A_{6,3}, \mathcal A_{6,4}, \mathcal A_{6,14}^{-1}$ plus the already studied cases of $\mathcal A_{3,1}$ and $\mathcal A_{6,5}^{\a}$. 
Let us now proceed to their geometric data.

\paragraph{${\cal A}_{5,1}$.} Let us first note that this case was also studied in \cite{kstt}. 
The commutation relations of the algebra are
\be 
[X_3,X_5]=X_1, \quad [X_4,X_5]=X_2.
\ee Then we find the invariant 1-forms
\bea 
e^1=dx^1-x^3dx^5, \quad e^2=dx^2-x^4dx^5, \quad e^i=dx^i, \quad i=3,4,5,
\eea 
which determine the twist matrix $e^a=U(x)^a_{b}dx^{b}$,
\be
U=\begin{pmatrix}
	1&0&0&0&-x^3 \\
	0&1&0&0&-x^4 \\
	0&0&1&0&0 \\
	0&0&0&1&0 \\
	0&0&0&0&1 
  \end{pmatrix}.
\ee
The identification conditions for the compactification are 
\bea\label{id51} (x^1,x^2,x^3,x^4,x^5)&\sim&(x^1+a,x^2,x^3,x^4,x^5) \nn \\
&\sim&(x^1,x^2+a,x^3,x^4,x^5) \nn\\ &\sim&(x^1+ax^5,x^2,x^3+a,x^4,x^5) \nn \\ 
&\sim&(x^1,x^2+ax^5,x^3,x^4+a,x^5) \nn \\ &\sim&(x^1,x^2,x^3,x^4,x^5+a), \quad a\in\Z.
\eea

\paragraph{${\cal A}_{5,4}$.} The commutation relations in the present case are
\be [X_2,X_4]=X_1, \quad [X_3,X_5]=X_1. \ee The corresponding 1-forms are found to be
\bea 
e^1=dx^1-x^2dx^4-x^3dx^5, \quad e^i=dx^i, \quad i=2,3,4,5,
\eea
and the twist matrix 
\be 
U=\begin{pmatrix}
	1&0&0&-x^2&-x^3 \\
	0&1&0&0&0 \\
	0&0&1&0&0 \\
	0&0&0&1&0 \\
	0&0&0&0&1 
  \end{pmatrix}.
\ee
The identification conditions for the compactification are 
\bea\label{id54} (x^1,x^2,x^3,x^4,x^5)&\sim&(x^1+a,x^2,x^3,x^4,x^5) \nn \\
&\sim&(x^1+ax^4,x^2+a,x^3,x^4,x^5) \nn\\ &\sim&(x^1+ax^5,x^2,x^3+a,x^4,x^5) \nn \\ 
&\sim&(x^1,x^2,x^3,x^4+a,x^5) \nn \\ &\sim&(x^1,x^2,x^3,x^4,x^5+a), \quad a\in\Z.
\eea

\paragraph{${\cal A}_{6,3}$.} The commutation relations are given as 
\be 
[X_1,X_2]=iX_6, \quad [X_1,X_3]=iX_4, \quad [X_2,X_3]=iX_5,
\ee
leading to the 1-forms
\bea
e^1&=&dx^1, \quad e^2=dx^2, \quad e^3=dx^3, \nn\\ \quad e^4&=&dx^4-x^1dx^3, \quad e^5=dx^5-x^2dx^3, \quad e^6=dx^6-x^1dx^2.
\eea
Thus the twist matrix turns out to be
\be 
U=\begin{pmatrix}
	1&0&0&0&0&0 \\
	0&1&0&0&0&0 \\
	0&0&1&0&0&0 \\
	0&0&-x^1&1&0&0 \\
	0&0&-x^2&0&1&0 \\
	0&-x^1&0&0&0&1 
  \end{pmatrix},
\ee
and the twisted identifications are 
\bea (x^1,x^2,x^3,x^4,x^5,x^6)&\sim&(x^1,x^2,x^3+a,x^4,x^5,x^6) \nn \\
&\sim&(x^1,x^2,x^3,x^4+a,x^5,x^6) \nn\\ &\sim&(x^1,x^2,x^3,x^4,x^5+a,x^6) \nn \\ 
&\sim&(x^1,x^2,x^3,x^4,x^5,x^6+a) \nn\\ &\sim&(x^1+a,x^2,x^3,x^4+ax^3,x^5,x^6+ax^2)\nn\\
&\sim&(x^1,x^2+a,x^3,x^4,x^5+ax^3,x^6), \quad a\in\Z.
\eea

\paragraph{${\cal A}_{6,4}$.} The commutation relations in the present case are
\be 
[X_1,X_2]=X_5, \quad [X_1,X_3]=X_6, \quad [X_2,X_4]=X_6.
\ee Then we find
\bea 
e^5=dx^5-x^1dx^2, \quad e^6=dx^6-x^1dx^3-x^2dx^4, \quad e^i=dx^i, \quad i=1,2,3,4,
\eea
and the twist matrix
\be 
U=\begin{pmatrix}
	1&0&0&0&0&0 \\
	0&1&0&0&0&0 \\
	0&0&1&0&0&0 \\
	0&0&0&1&0&0 \\
	0&-x^1&0&0&1&0 \\
	0&0&-x^1&-x^2&0&1 
  \end{pmatrix},
\ee
The identification conditions for the compactification are 
\bea\label{id64} (x^1,x^2,x^3,x^4,x^5,x^6)&\sim&(x^1+a,x^2,x^3,x^4,x^5+ax^2,x^6+ax^3) \nn \\
&\sim&(x^1,x^2+a,x^3,x^4,x^5,x^6+ax^4) \nn\\ &\sim&(x^1,x^2,x^3+a,x^4,x^5,x^6) \nn \\ 
&\sim&(x^1,x^2,x^3,x^4+a,x^5,x^6) \nn \\ &\sim&(x^1,x^2,x^3,x^4,x^5+a,x^6)\nn\\
&\sim&(x^1,x^2,x^3,x^4,x^5,x^6+a), \quad a\in\Z.
\eea

\paragraph{${\cal A}_{6,14}^{-1}$.} The Lie algebra commutation relations are
\be 
[X_1,X_3]=X_4, \quad [X_1,X_4]=X_6, \quad [X_2,X_3]=X_5, \quad [X_2,X_5]=-X_6.
\ee 
The invariant 1-forms are found to be
\bea 
 e^i&=&dx^i, \quad i=1,2,3, \nn\\ 
e^4&=&dx^4-x^1dx^3, \quad e^5=dx^5-x^2dx^3, \nn\\ e^6&=&dx^6-x^1dx^4+((x^1)^2+(x^2)^2)dx^3+(x^2x^3-x^5)dx^2,
\eea
while the following additional relations are obtained
\bea
x^1dx^1+x^2dx^2=0 \quad &\Leftrightarrow &\quad (x^1)^2+(x^2)^2= \mbox{const.}, \nn\\
x^2dx^3+x^3dx^2=0 \quad &\Leftrightarrow &\quad x^2x^3=\mbox{const.}
\eea
The twist matrix is
\be 
U=\begin{pmatrix}
	1&0&0&0&0&0 \\
	0&1&0&0&0&0 \\
	0&0&1&0&0&0 \\
	0&0&-x^1&1&0&0 \\
	0&0&-x^2&0&1&0 \\
	0&x^2x^3-x^5&(x^1)^2+(x^2)^2&-x^1&0&1 
  \end{pmatrix},
\ee
and determining the identification conditions for the compactification in the present case 
turns out to be complicated due to the $x^5$ dependence of $U$.


\begin{thebibliography}{99}


\bibitem{Banks:1996vh}

  T.~Banks, W.~Fischler, S.~H.~Shenker and L.~Susskind,
  Phys.\ Rev.\  {\bf D55 } (1997)  5112
  [hep-th/9610043].


\bibitem{Ishibashi:1996xs}

 N.~Ishibashi, H.~Kawai, Y.~Kitazawa and A.~Tsuchiya,
 Nucl.\ Phys.\  B {\bf 498} (1997) 467
 [arXiv:hep-th/9612115].


\bibitem{Taylor:1996ik}
  W.~Taylor,
  Phys.\ Lett.\ B {\bf 394} (1997) 283
  [hep-th/9611042].



\bibitem{cds}
  A.~Connes, M.~R.~Douglas and A.~S.~Schwarz,\\
  JHEP {\bf 9802 } (1998)  003 
  [hep-th/9711162].

\bibitem{Douglas:1997fm}
  M.~R.~Douglas and C.~M.~Hull,
  JHEP {\bf 9802} (1998) 008
  [hep-th/9711165].



\bibitem{Brace:1998ku}

  D.~Brace, B.~Morariu and B.~Zumino,  
  Nucl.\ Phys.\ B {\bf 545} (1999) 192
  [hep-th/9810099].


\bibitem{Kawano:1998re}
  T.~Kawano and K.~Okuyama,
  Phys.\ Lett.\ B {\bf 433} (1998) 29
  [hep-th/9803044].


\bibitem{ramgoolam}
  D.~A.~Lowe, H.~Nastase and S.~Ramgoolam,\\
  Nucl.\ Phys.\ B {\bf 667} (2003) 55
  [hep-th/0303173].




\bibitem{Ho:1997yk}
  P.~-M.~Ho, Y.~-Y.~Wu and Y.~-S.~Wu,
  Phys.\ Rev.\ D {\bf 58} (1998) 026006
  [hep-th/9712201].

\bibitem{Ho:1998xh}
  P.~-M.~Ho and Y.~-S.~Wu,
  Phys.\ Rev.\ D {\bf 58} (1998) 066003
  [hep-th/9801147].

\bibitem{Chatzistavrakidislie}
  A.~Chatzistavrakidis,
  Phys.\ Rev.\ D {\bf 84} (2011) 106010
  [arXiv:1108.1107 [hep-th]].

\bibitem{Nishimura}
  S.~-W.~Kim, J.~Nishimura and A.~Tsuchiya,
``Expanding universe as a classical solution in the Lorentzian matrix model for nonperturbative superstring theory'',
  arXiv:1110.4803 [hep-th].


\bibitem{km}
  N.~Kaloper and R.~C.~Myers,
  JHEP {\bf 9905} (1999) 010
  [hep-th/9901045].

\bibitem{Hull:2005hk}
  C.~M.~Hull and R.~A.~Reid-Edwards,
  Fortsch.\ Phys.\  {\bf 57} (2009) 862
  [hep-th/0503114].

\bibitem{Hull:2006tp}
  C.~M.~Hull and R.~A.~Reid-Edwards,
  JHEP {\bf 0610} (2006) 086
  [hep-th/0603094].

\bibitem{SW}
N.~Seiberg and E.~Witten,
JHEP  {\bf 9909}, 032 (1999) [arXiv:hep-th/9908142].



\bibitem{Scherk}
  J.~Scherk and J.~H.~Schwarz,
  Nucl.\ Phys.\ B {\bf 153} (1979) 61.






\bibitem{kstt}
  S.~Kachru, M.~B.~Schulz, P.~K.~Tripathy and S.~P.~Trivedi,
  JHEP {\bf 0303} (2003) 061
  [hep-th/0211182].



\bibitem{Grana}

  M.~Grana, R.~Minasian, M.~Petrini and A.~Tomasiello,\\
  JHEP {\bf 0705 } (2007)  031
  [hep-th/0609124].






\bibitem{malcev}

A.~I.~Mal'cev, 
AMS Translation No. 39 (1951).






\bibitem{Patera:1976ud}
  J.~Patera, R.~T.~Sharp, P.~Winternitz and H.~Zassenhaus,
  J.\ Math.\ Phys.\ (N.Y.)  {\bf 17 } (1976)  986.


\bibitem{Buscher:1987qj}
  T.~H.~Buscher,
  Phys.\ Lett.\ B {\bf 201} (1988) 466.



\bibitem{Madore:2000en}
  J.~Madore, S.~Schraml, P.~Schupp and J.~Wess,
  Eur.\ Phys.\ J.\  {\bf C16 } (2000)  161-167.
  [hep-th/0001203].

\bibitem{Connes:1994yd}
  A.~Connes,
  ``Noncommutative geometry'', Academic Press, San Diego, CA, 1994.

\bibitem{Madore:2000aq}
  J.~Madore,
  ``An introduction to noncommutative differential geometry and its physical applications'',
  Lond.\ Math.\ Soc.\ Lect.\ Note Ser.\  {\bf 257 }, London Mathematical Society, London, 2000.


\bibitem{Anazawa:1999ub}
  M.~Anazawa,
  Nucl.\ Phys.\ B {\bf 569} (2000) 680
  [hep-th/9905055].


\bibitem{Lust:2010iy}
  D.~L\"ust,
  JHEP {\bf 1012} (2010) 084
  [arXiv:1010.1361 [hep-th]].

\bibitem{Blumenhagen:2010hj}
  R.~Blumenhagen and E.~Plauschinn,\\
  J.\ Phys.\  A {\bf 44} (2011) 015401
  [arXiv:1010.1263 [hep-th]].


\bibitem{Blumenhagen:2011ph}
  R.~Blumenhagen, A.~Deser, D.~L\"ust, E.~Plauschinn and F.~Rennecke,
  J.\ Phys.\ A  {\bf 44} (2011) 385401
  [arXiv:1106.0316 [hep-th]].

\bibitem{Cornalba:2001sm}
  L.~Cornalba and R.~Schiappa,
  Commun.\ Math.\ Phys.\  {\bf 225} (2002) 33
  [hep-th/0101219].

\bibitem{Ramgoolam:2003cs}
  S.~Ramgoolam,
  JHEP {\bf 0403} (2004) 034
  [hep-th/0310153].

\bibitem{Brace:1998xz}
  D.~Brace, B.~Morariu and B.~Zumino,
  Nucl.\ Phys.\ B {\bf 549} (1999) 181
  [hep-th/9811213].




\bibitem{Ganor:1996zk}
  O.~J.~Ganor, S.~Ramgoolam and W.~Taylor,\\
  Nucl.\ Phys.\ B {\bf 492} (1997) 191
  [hep-th/9611202].

\bibitem{Hull:2004in}
  C.~M.~Hull,
  JHEP {\bf 0510} (2005) 065
  [hep-th/0406102].



\bibitem{Grange:2006es}
  P.~Grange and S.~Sch\"afer-Nameki,
  Nucl.\ Phys.\ B {\bf 770} (2007) 123
  [hep-th/0609084].

\bibitem{Andriot:2011uh}
  D.~Andriot, M.~Larfors, D.~L\"{u}st and P.~Patalong,
  JHEP {\bf 1109} (2011) 134 
  [arXiv:1106.4015 [hep-th]].














\end{thebibliography}
\end{document}